\documentclass[prl,twocolumn,longbibliography]{revtex4-1}
\usepackage{amsmath}
\usepackage{amssymb}
\usepackage{natbib}
\usepackage{subfigure}
\usepackage{mathtools}
\usepackage{graphicx}
\usepackage{color}
\usepackage{setspace}
\usepackage{mathrsfs}
\usepackage[usenames,dvipsnames]{xcolor}
\newcommand{\beginsupplement}{%
        \setcounter{table}{0}
        \renewcommand{\thetable}{S\arabic{table}}%
        \setcounter{figure}{0}
        \renewcommand{\thefigure}{S\arabic{figure}}%
     }

\def\bdot{\raise.2em\hbox to .15em{.}}

\definecolor{gray}{gray}{0.5}

\def\bdotblack{\raise.25em\hbox to .15em{.}}

\definecolor{pinegreen}{rgb}{0.0, 0.47, 0.44}
\definecolor{debianred}{rgb}{0.84, 0.04, 0.33}

\begin{document}
\title{Revealing the state space of turbulence using machine learning}
\author{Jacob Page$^{1,2}$}
\author{Michael P. Brenner$^{3,4}$}
\author{Rich R. Kerswell$^2$}
%
%
\affiliation{$^{1}$School of Mathematics, University of Edinburgh, Edinburgh, EH9 3FD, UK}
\affiliation{$^{2}$DAMTP, Centre for Mathematical Sciences, University of Cambridge, Cambridge, CB3 0WA, UK}
\affiliation{$^{3}$School of Engineering and Applied Sciences, Harvard University, Cambridge MA 02138}
\affiliation{$^{4}$Google Research, Mountain View, CA 94043}

\begin{abstract}
Despite the apparent complexity of turbulent flow, identifying a simpler description of the underlying dynamical system remains a fundamental challenge. Capturing how the turbulent flow meanders amongst unstable states (simple invariant solutions) in phase space, as envisaged by Hopf in 1948, using some efficient representation offers the best hope of doing this, despite the inherent difficulty in identifying these states. Here, we make a significant step towards this goal by demonstrating that deep convolutional autoencoders can identify low-dimensional representations of two-dimensional turbulence which are closely associated with the simple invariant solutions characterizing the turbulent attractor. To establish this, we develop \emph{latent Fourier analysis} that decomposes the flow embedding into a set of orthogonal latent Fourier modes which decode into physically meaningful patterns resembling simple invariant solutions.  
The utility of this approach is highlighted by analysing turbulent Kolmogorov flow (flow on a 2D torus forced at large scale) at $Re=40$ where, in between intermittent bursts, the flow resides in the neighbourhood of an unstable state and is very low dimensional. Projections onto individual latent Fourier wavenumbers reveal the simple invariant solutions organising both the quiescent and  bursting dynamics in a systematic way inaccessible to previous approaches.

\end{abstract}

\maketitle


Building effective low-order representations of turbulent flows is a long-standing challenge that could dramatically improve our capabilities for prediction and control. 
Current state-of-the-art techniques for low-order modelling typically involve constructing a set of orthogonal `modes' from a dataset.
Perhaps most well known is principal component analysis (PCA), which produces an orthogonal basis to optimally represent the training snapshots. 
However, while highly interpretable, modes in the basis may have little dynamical significance individually \cite{Rowley2017},  and other methods that attempt to also infer dynamics -- for example dynamic mode decomposition  \cite{Schmid2010} -- are ill-suited to chaotic systems like turbulence  \cite{Page2019}.
The failure of these low-order representations to faithfully reconstruct even weak turbulence contrasts with the dynamical systems view of the flow, in which turbulence is understood to arise as the structure of phase state complexifies under increasing Reynolds number, $Re$ \cite{Landau1944,Hopf1948,Kerswell2005,Eckhardt2007,Kawahara2012}.
In this framework, a turbulent flow is considered as a long nonclosing orbit in a high-dimensional state space, transiting between unstable simple invariant solutions which are the `building blocks' of the chaotic attractor \cite{Hopf1948}. Such a viewpoint suggests that there are efficient low-order representations of the flow which are rooted in the underlying simple invariant solutions,
though the nonlinearity of the Navier-Stokes equations confounds our attempts to hand-craft a solution.

The recent emergence of deep convolutional neural networks (CNNs) represents an opportunity to identify such representations due to their ability to extract patterns \cite{LeCun2015,Gulshan2016} that can result in highly efficient low-dimensional embeddings of complex data.
The utility of CNNs in the study of nonlinear partial differential equations (PDEs) has been demonstrated recently in a number of canonical examples,
where their accurate paramterisation of the solution manifold has been exploited to successfully predict chaotic dynamics for multiple Lyapunov times \cite{Pathak2018}, to estimate the dimension of chaotic attractors \cite{Linot2020} and to design new spatial discretisation schemes \cite{Bar-Sinai2019}.

Using a CNN to decompose a turbulent flow into a series of recurrent spatial \emph{patterns} should be contrasted to a projection onto a hand-crafted orthogonal basis such as Fourier modes, 
where the coupling of all wavenumbers through the nonlinearity of the Navier-Stokes equation renders individual modes dynamically insignificant.  
A learnt basis  has the potential to encode and parameterise the alphabet of dynamical processes present, though at a loss of physical interpretability.
%
%
Here we show how the presence of a continuous symmetry in the physical system can be exploited to perform a decomposition of embeddings of a turbulent flow in latent space. This \emph{latent} Fourier analysis -- analogous to a Fourier decomposition in physical space -- 
yields a (latently) orthogonal basis of recurrent patterns that exhibit striking resemblance to simple invariant solutions of the underlying dynamical system. 

\section*{Dimensionality reduction of Kolmogorov flow}

We use deep CNNs to build efficient low-dimensional representations of snapshots from a computation of monochromatically forced, two-dimensional turbulence on a doubly-periodic square $[0,2\pi]\times [0,2\pi]$.
In two dimensions the Navier-Stokes equations can be combined and written concisely in terms of the out-of-plane vorticity 
$\omega := (\boldsymbol \nabla \times \mathbf u)\cdot \hat{\mathbf z}$, 
\begin{equation}
    \partial_t\omega + \mathbf u\cdot\boldsymbol \nabla\omega = \frac{1}{Re}\nabla^2 \omega + (\boldsymbol \nabla \times \mathbf f)\cdot \hat{\mathbf z},
    \label{eqn:NS}
\end{equation}
where $\mathbf f = \sin 4y \,\hat{\mathbf x}$ (`Kolmogorov' flow \cite{Arnold1960} with the specific choice of four forcing wavelengths in the square \cite{Platt1991,Chandler2013,Farazmand2016}). 
There are a number of symmetries,
the most important in the context of this work being the continuous translational symmetry $\mathscr T_{s}: \omega(x,y) \to \omega(x+s,y)$.
There is also a discrete shift-reflect symmetry $\mathscr S: \omega(x,y) \to -\omega(-x, y+\pi/4)$
and a rotational symmetry $\mathscr R: \omega(x,y) \to \omega(-x, -y)$.
Throughout we hold the Reynolds number fixed at $Re=40$, where a large number of simple invariant solutions have been found \cite{Chandler2013,Farazmand2016}. 

We seek efficient low-dimensional \emph{embeddings} $\{\pmb{\mathscr E}\}$ of vorticity snapshots $\{\omega\}$, which are essentially greyscale images of dimension $N_x\times N_y = 128 \times 128$.
To do this we construct deep CNNs in the form of autoencoders, which are trained to reconstruct the input snapshots with dimension reduction applied as part of the network structure.
The specific architectures we use are described in detail in the Supplementary Information (SI), but all consist of an encoder module,
$\pmb{\mathscr E} \!\!:\mathbb R^{128\times 128}\to \mathbb R^m$ -- consisting of
a series of five convolutional layers with pooling to reduce the dimension from $128^2$ to $128$ -- followed by fully connected layers to further reduce the dimension to $m \leq 128$ (the ``embedding''). 
A similar structure is used to decode the embeddings, $\mathscr D:\mathbb R^m \to \mathbb R^{128\times 128}$, 
and the weights that define these functions are obtained by performing stochastic gradient descent on the loss functional $\tfrac{1}{|\text{data}|}\sum_{\text{data}}\|[\mathscr D\circ \pmb{\mathscr E}](\omega) - \omega\|^2_2$.

%
%
%
\begin{figure*}
    \includegraphics[width=\textwidth]{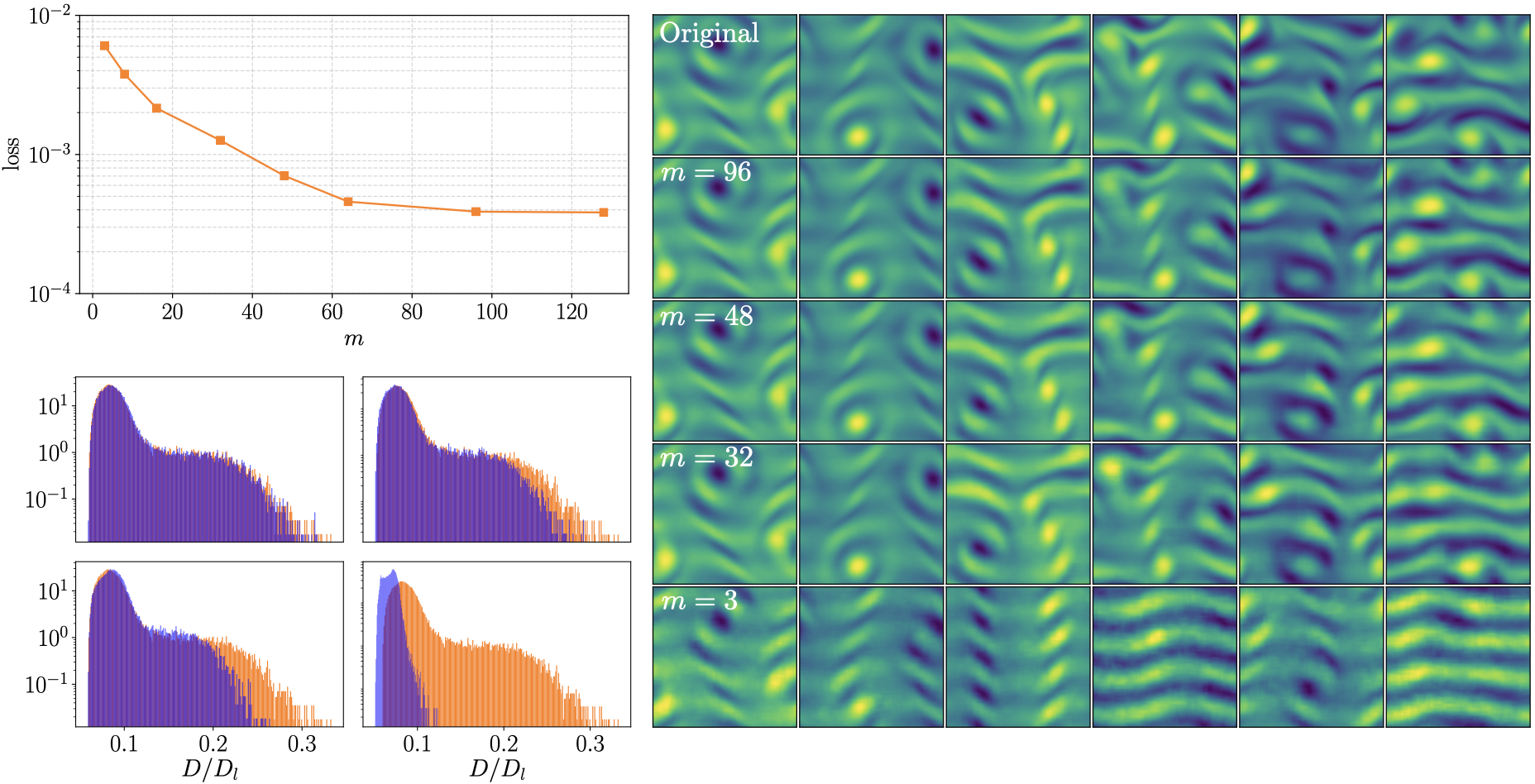}
    \caption{
    \textbf{Autoencoders for dimensionality reduction in Kolmogorov flow.}
    (Top left) Loss $\tfrac{1}{|\text{data}|}\sum_{\text{data}}\|[\mathscr D\circ \pmb{\mathscr E}](\omega) - \omega\|^2_2$ as a function of embedding dimension $m$.
    (Bottom left) Dissipation PDFs of autoencoded vorticity fields. The background orange is obtained from the original test dataset, the overlayed blue PDFs are for $m\in\{96,48,32,3\}$ (top left to bottom right).
    (Right) Decodes of a set of snapshots of increasing dissipation for various embedding dimensions $m$. 
    Snapshots are ordered left-to-right by dissipation, running from $D/D_l\approx 0.08$ to $D/D_l\approx 0.26$ indicating, for example, that $m=3$ captures low-dissipation episodes well but struggles to represent high-dissipation events. 
    }
    \label{fig:loss}
\end{figure*}
The impact of autoencoder dimension on the fidelity of the reconstruction $[\mathscr D\circ\pmb{\mathscr E}](\omega)$ is examined in figure \ref{fig:loss}.
The loss drops monotonically with increasing $m$, with even very low dimensional networks (e.g. $m=3$) displaying relatively small losses (two unrelated vorticity fields typically yield an $O(1)$ loss), which suggests that much of the underlying dynamical system may be 
low dimensional. 
Furthermore, networks with modest $m$ (e.g. see the PDF for $m=32$ in figure \ref{fig:loss}) retain much of the high-dissipation tail in PDFs of   
$D := \langle (\boldsymbol \nabla \mathbf u)^2\rangle_V =\langle \omega^2 \rangle_V$. 
This indicates a retention of smaller scales and sharp variations in $\omega$ under dimensionality reduction, in contrast to standard techniques like PCA.
Even at $m=3$, the accurate reproductions of low dissipation events (see the snapshots in figure \ref{fig:loss}) contain the full spectrum of Fourier modes. 

To examine how these networks can reduce the dimensionality of the data while retaining a broad spectrum of lengthscales, we
describe a method for decomposing the latent representations of vortical snapshots into a finite set of recurrent patterns which can be visualised individually.

\section*{Latent Fourier analysis}
%
%
\begin{figure*}
    \centering
    \includegraphics[width=0.95\textwidth]{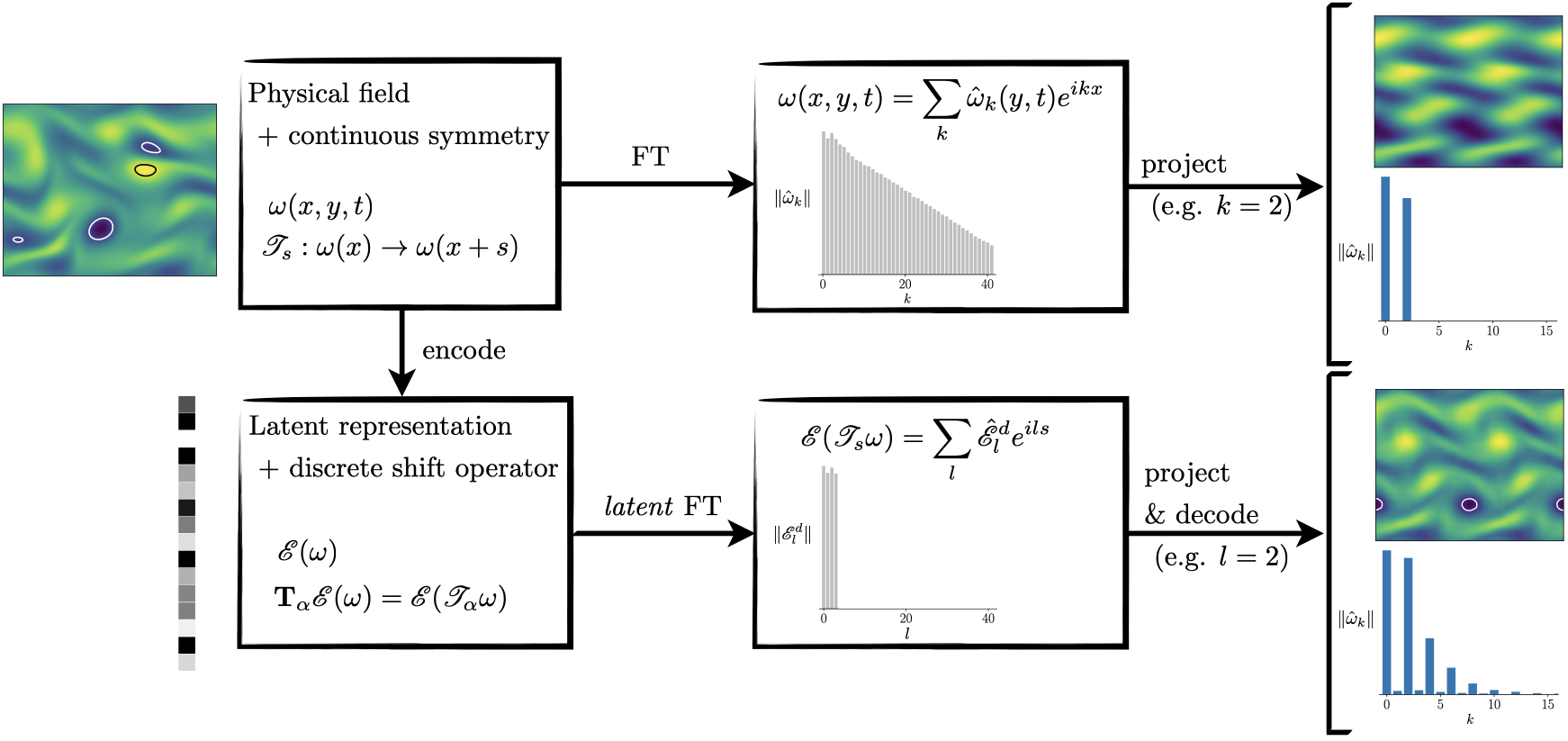}
    \caption{
    \textbf{An overview of the approach to performing latent Fourier decompositions on (latent) representations of
    physical systems with a continous symmetry.}
    A latent Fourier decomposition is performed by constructing an operator to map between embeddings of shifted versions of the same snapshot. 
    Note that only four latent wavenumbers, $\{0,1,2,3\}$, are required for the monochromatically forced turbulence considered here. 
    In the ``project and decode'' step (bottom right) the projection within the $l=0$ subspace is also included (see the discussion in the text).
    For comparison, the projection onto $k=0$ is also included in the projection onto the physical Fourier modes.
    }
    \label{fig:schem}
\end{figure*}
The continuous translational symmetry in the governing equation (\ref{eqn:NS}) and boundary conditions provides a mechanism to decompose the latent embeddings of the vorticity fields into recurrent `patterns' that reveal the structure of the latent space.
This decomposition is analogous to a Fourier transform in physical space, $\omega(x,y,t) = \sum_{k\in \mathbb Z} \hat{\omega}_k(y,t)\text{exp}(i k x)$.
However, the autoencoder representations encode horizontal position $x$  in an unknown way. To perform a similar decomposition for embeddings, we first  must construct an operator that can map between an embedding of a snapshot and an embedding of a shifted version of the same field:
$\mathbf T_{\alpha}\pmb{\mathscr E}(\omega) = \pmb{\mathscr E}(\mathscr T_{\alpha}\omega)$,
where the fixed shift $\alpha \in (0,2\pi)$ is a design choice.
This procedure applied to the vorticity field itself would result in a numerical approximation to a Fourier transform through the eigenvalues and eigenvectors of $\mathbf T_{\alpha}$, with a maximum resolved wavenumber set by the Nyquist condition, $k_{max} = \pi /\alpha$.
For the embeddings, $\pmb{\mathscr E}$, the value of $\alpha$ sets a maximum \emph{latent} wavenumber that can be resolved, $l_{max}$.
As we will show below, the required latent resolution is considerably coarser than the smallest scales generated by the governing PDE (\ref{eqn:NS}).

To see the connection to a standard Fourier transform, consider a discrete shift $\alpha = 2\pi /n$, with $n\in \mathbb N$.
By design, $\mathbf T_{\alpha}^n\pmb{\mathscr E}(\omega) = \pmb{\mathscr E}(\omega)$, so the eigenvalues of $\mathbf T_{\alpha}$ 
are $\Lambda_j=\text{exp}(2\pi i l_j/n)$, with $l_j\in \mathbb Z$. 
We assume that the value of $\alpha$ has been chosen small enough so that no $l_j$ are found beyond a maximum ($l_{max}$).
With this, approximations to continuous shifts $s$ of an embedding $\pmb{\mathscr E}(\omega)$ are 
\begin{align}
    \pmb{\mathscr E}(\mathscr T_s\omega) &= 
    \sum_{l}\left( \sum_{j=0}^{d(l)-1} 
    \pmb{\mathcal P}^l_j (\pmb{\mathscr E}(\omega))\right) e^{ils} \nonumber \\ 
    &:=
    \sum_{l}\left(\sum_{j=0}^{d(l)-1} 
    [(\boldsymbol \xi_j^{\dagger (l)})^H \pmb{\mathscr E}(\omega)] \boldsymbol \xi_j^{(l)}
    \right) e^{ils}.
    \label{eqn:latent_FT}
\end{align}
Here $l$ is the \emph{latent wavenumber} 
and the operator $\pmb{\mathcal P}_j^l$ is a projector in direction $j$ within the eigenspace of wavenumber $l$, which has geometric multiplicity $d(l)$. 
Unlike physical Fourier modes, the latent wavenumbers are degenerate.
Equation (\ref{eqn:latent_FT}) assumes that some bi-orthogonal basis has been constructed; a specific choice is discussed further below.

The number of required latent wavenumbers and their degeneracy provides insight into the nonlinear interactions in physical space.
Each latent Fourier mode of wavenumber $l$ can be decoded into a $2\pi/l$-periodic pattern which has a physical Fourier decomposition projecting onto wavenumbers $k_q = ql$, $q \in \mathbb N$ (see the example in the schematic of figure \ref{fig:schem}).
These recurrent patterns represent pathways through physical Fourier wavenumbers which are selected by the dynamics.

%
%
\begin{figure}
    \centering
    \includegraphics[width=0.48\textwidth]{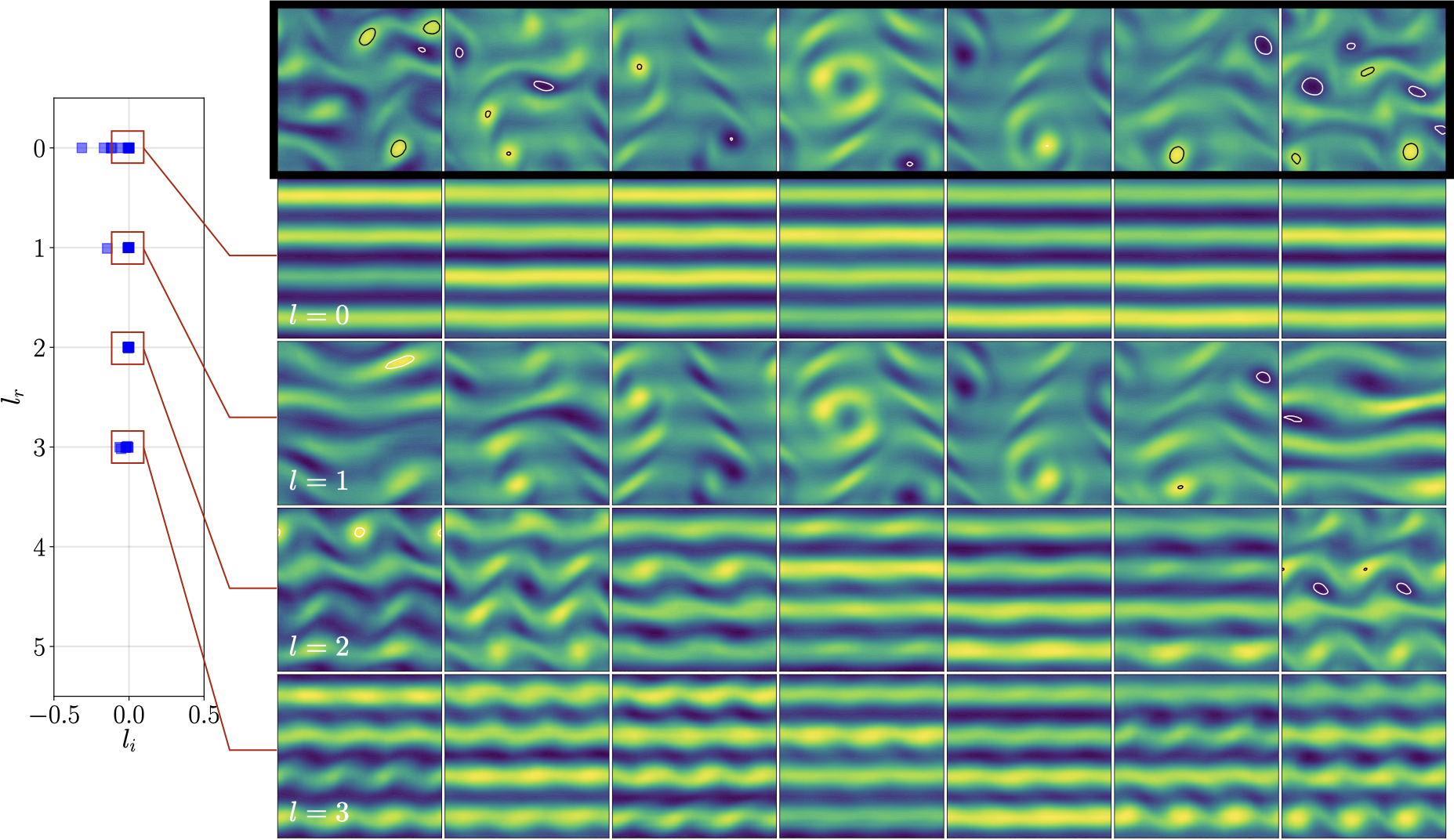}
    \caption{\textbf{Eigenvalue spectra of a latent symmetry operator and decodes of the projection onto individual eigenspaces for example snapshots.} 
    (Left) Eigenvalue spectrum of $\mathbf T_{\alpha}$ on the $m=96$ autoencoder with $\alpha = 2\pi/9$. Note that only half the spectrum is shown.
    (Right) Seven example snapshots (top row, inside black box) and the decodes of the projection of their embeddings onto individual latent Fourier eigenspaces.
    Note that the $l=0$ subspace is always included for decodes of latent wavenumbers $l>0$.}
    \label{fig:Tspec}
\end{figure}
To perform a latent Fourier decomposition within our autoencoder, we build a shift operator using a least-squares fit to find a $\mathbf T_{\alpha}$ that maps between embeddings of the test set and embeddings of the same vorticity fields shifted by $\alpha$ in $x$ 
(see SI).
Numerical experiments reducing $\alpha$ reveal a maximum latent wavenumber $l_{max}=3$ for embeddings $m\geq 32$. 
This truncation suggests that the learnt representations can be decomposed into a set of recurrent patterns which are at most $2\pi/3$-periodic.
The energy in all higher physical wavenumbers, $k>3$, is assigned during the decode of this coarse set of features in the latent space (e.g. $k=5$ can only be encoded into $l=1$ and $k=8$ into  $l=2$).  An example eigenvalue spectrum for the $m=96$ network is reported in figure \ref{fig:Tspec}, the (degenerate) eigenvalues lying approximately on 
$l\in\{0, 1, 2, 3\}$. 
Some example $2\pi/l$-periodic physical patterns associated with a particular value of $l$ are displayed alongside the spectrum.
These images were generated by projecting the embedding of a snapshot onto the relevant eigenspace and decoding the result. 
Note that, in contrast to a standard Fourier transform, the $l=0$ contribution must always be included for the decode operation to yield a physical field. 
Projections onto the $l=0$ subspace decode to horizontal stripes of vorticity which align with the (curl of) the forcing in equation (\ref{eqn:NS}); the vorticity amplitude of each stripe is distorted by the $l>0$ modes into vortical features. 
In this way, much of the $y$-dependence in the final decoded snapshot is controlled by the $l=0$ subspace. 

The decodes in figure \ref{fig:Tspec} for $l>0$  have the expected periodicity, $\mathscr T_{2\pi/l}[\mathscr D(\sum_j \pmb{\mathcal P}_j^l(\pmb{\mathscr E}))] = \mathscr D(\sum_j \pmb{\mathcal P}_j^l(\pmb{\mathscr E}))$. 
In contrast to a projection onto individual Fourier modes in physical space, the projection onto individual latent wavenumbers produces
patterns with vortical features that can be clearly identified in the original snapshots. 
The wide range of features observed in decodes of individual latent wavenumbers is possible due to the degeneracy of the eigenspaces, which we now discuss.  

\section*{Connections to simple invariant solutions}
%
%
\begin{figure}
    \centering
    \includegraphics[width=0.48\textwidth]{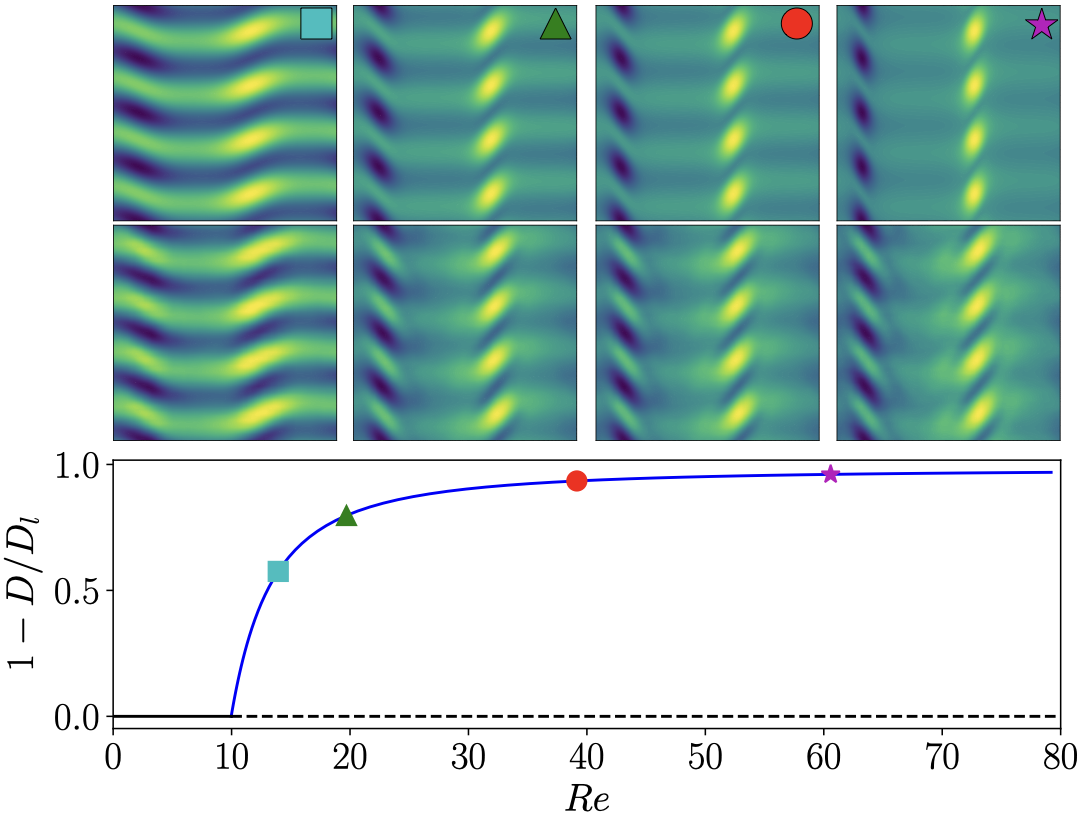}
    \caption{\textbf{The primary bifurcation of Kolmogorov flow and its connection to the primary latent Fourier mode within the $l=1$ subspace.}
    (Top) Equilibrium states at $Re\approx\{14, 20, 40, 60\}$ from the solution branch which bifurcates from the basic laminar state at $Re \approx 9.97$.
    (Middle) Decode of the projection of the embeddings of these equilibria onto the first PCA modes within the $l=0$ and $l=1$ subspaces.
    (Bottom) Amplitude (measured by disspation) of this equilibrium as a function of Reynolds number (blue line, the laminar solution sits on $1-D/D_l=0$)  with the example states shown above identified with symbols. 
    It should be emphasised that training has been conducted at fixed $Re=40$, and that all of the training snapshots are from within the turbulent attractor and do not feature this simple equilibrium.}
    \label{fig:bifurcation}
\end{figure}

The degenerate set of recurrent patterns encoded within each latent eigenspace can be revealed by an appropriate choice of basis to define the projectors $\pmb{\mathcal P}_j^l$ in equation (\ref{eqn:latent_FT}).
We have found PCA within each eigenspace to be robust for this purpose. 
The decomposition within $l=0$, while not particularly informative on its own, is most useful for visualising other eigenspaces,
because a projection onto the leading PCA mode in $l=0$, $\boldsymbol u_0^{(l=0)}$, decodes a vorticity field resembling the laminar parallel flow solution.
This field is invariant under all symmetry operations, allowing symmetries within the $l>0$ eigenspaces to be identified. 

    A singular value decomposition within the $l=1$ subspace reveals the presence of a large-amplitude leading mode, with higher order modes appearing in pairs at lower energies (see figure \ref{fig:lat_pca} in the SI).
    For visualisation of individual $l=1$ modes we consider decodes of the projection with only the leading PCA mode from the $l=0$ subspace included,
    \begin{equation}
        \overline{\omega}_n^{(1)} = \mathscr D\left( \pmb{\mathcal P}_0^0 + [\pmb{\mathcal P}_n^1 + \text{c.c.}]\right).
        \label{eqn:l_pca_v2}
    \end{equation}
    As described above, the use of $\pmb{\mathcal P}_0^0$ alone removes much of the $y$-dependence when visualising $l=1$ modes from projections of arbitrary snapshots due to the high degree of symmetry associated with $\boldsymbol u_0^{(0)}$.  
    A specific example of this is included in figure \ref{fig:lat_pca} in the SI, and shows that the PCA modes within $l=1$ decode structures which also have a number of discrete symmetries. 

    The large-amplitude primary PCA mode in the $l=1$ subspace, $\boldsymbol u_0^{(1)}$, has a particular physical significance. 
    Decoding projections $\overline{\omega}_0^{(1)}$ (equation \ref{eqn:l_pca_v2}) reveal a structure that is symmetric under rotation and shift-reflects,
    $\overline{\omega}_0^{(1)} = \mathscr R \,\overline{\omega}_0^{(1)}$, $\overline{\omega}_0^{(1)}=\mathscr S^{m}\, \overline{\omega}_0^{(1)}$
    (see figure \ref{fig:lat_pca} in the SI)
    that strongly resembles the equilibrium born in the continuous-symmetry-breaking bifurcation off the laminar base state at low $Re\approx 10$ \cite{Chandler2013,Lucas2014}.
    We explore this connection in figure \ref{fig:bifurcation}, where we show that decodes of projections onto $\boldsymbol u_0^{(0)}$ and $\boldsymbol u_0^{(1)}$ can be used to reconstruct this structure over \emph{a range} of $Re$, despite the fact that the training was conducted at fixed $Re=40$.  
    As the solution branch is traversed, the amplitude of the projection of the embedding onto $\boldsymbol u_0^{(1)}$ is increased.
    In the vorticity field, this corresponds to both a strengthening and tilting of the vorticity bands. 

    While it is surprising that this non-trivial, $2\pi$-periodic equilibrium should form the backbone of the latent representations -- neither it nor the laminar solution are seen explicitly during training -- it is intuitive as further simple invariant solutions and the emergence of chaotic dynamics appear in bifurcations from this state.
It's worth emphasizing that this structure is associated with a \emph{single} latent wavenumber, $l=1$, in contrast to its physical Fourier transform which projects onto all physical wavenumbers.

%
%
\begin{figure}
    \centering
    \includegraphics[width=0.48\textwidth]{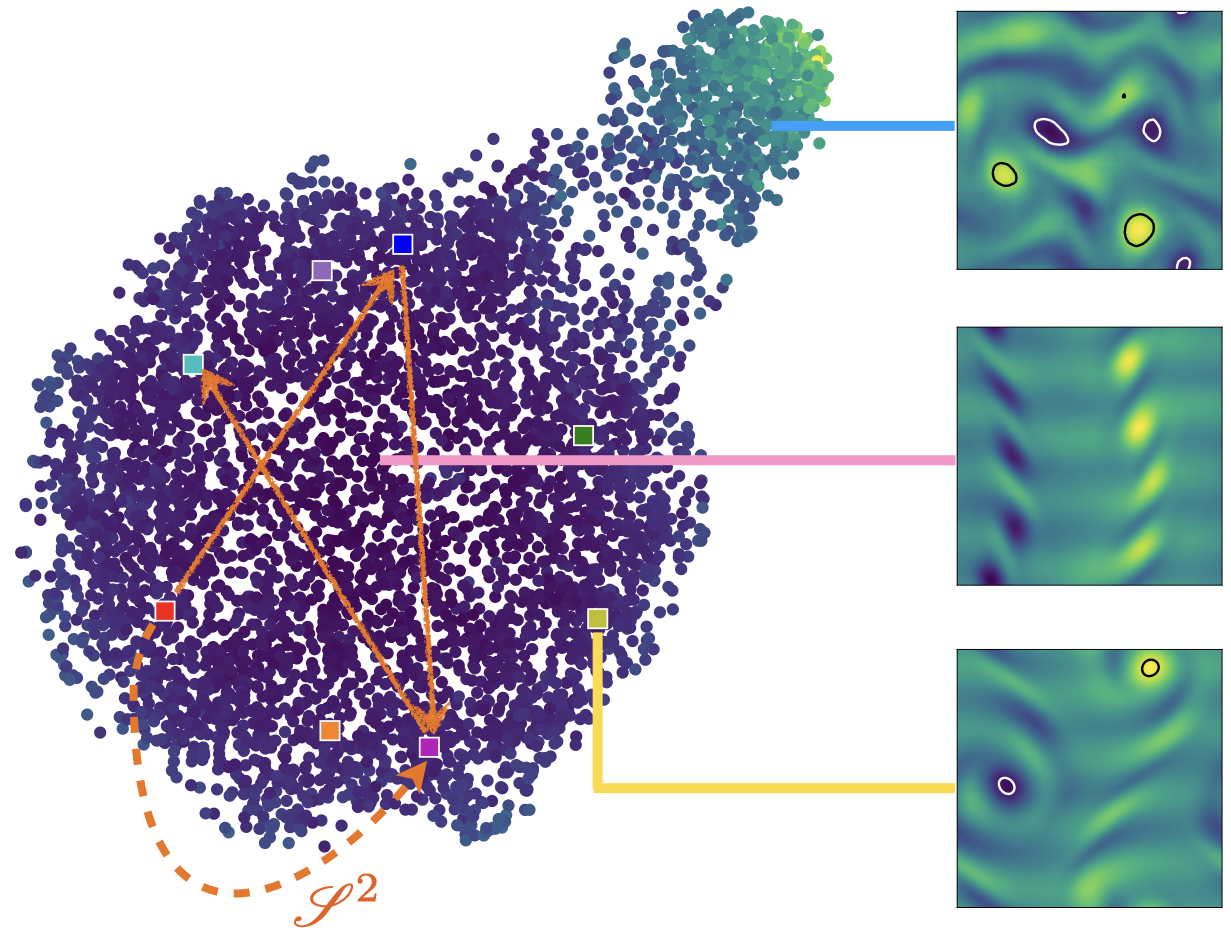}
    \caption{
        \textbf{Two-dimensional t-SNE of embeddings of 5000 vorticity snapshots in the $m=96$ network.}
        The figure was created by computing projections of the embeddings onto the five leading PCA modes within each degenerate eigenspace $l\in\{0,1,2,3\}$.
        Translational dependence of the features was then removed by taking the absolute value of these projections for $l>0$ 
        before the t-SNE algorithm\cite{tsne}
        was applied.
        The data points are coloured by their dissipation values which run between $D/D_l = 0.06$ (dark blue) and $D/D_l=0.33$ (yellow).
        Decodes of example snapshots are shown on the right.
        The eight coloured squares represent the embeddings of the same vorticity snapshot (see the decode of the yellow square at bottom right) with
        the shift-reflect operation repeatedly applied. 
        The orange arrows indicate how this field is moved between sectors of the low-dissipation octagon under applications of a shift-reflect operation; the dashed arrow labelled $\mathscr S^2$ corresponds to a full wavelength shift in $y$.
    }
    \label{fig:tsne}
\end{figure}


The full structure of the state space of vorticity fields can be concisely visualised by first projecting embeddings of the test dataset onto the latent Fourier modes:
\begin{equation}
    \boldsymbol \psi(\omega) := \begin{pmatrix} \boldsymbol u_0^{0H} \pmb{\mathscr{E}}(\omega) \\
        \boldsymbol u_1^{0H} \pmb{\mathscr{E}}(\omega) \\
        \vdots \\
        |\boldsymbol u_0^{1H} \pmb{\mathscr{E}}(\omega)| \\ 
        \vdots \\
        |\boldsymbol u_0^{2H} \pmb{\mathscr{E}}(\omega)| \\
        \vdots
    \end{pmatrix},
    \label{eqn:lat_obs}
\end{equation}
where the first five PCA modes of each eigenspace are included, and taking the absolute value of projections onto PCA modes from eigenspaces $l\geq 1$ removes any dependence on the relative streamwise location of the recurrent patterns.
A two-dimensional visualisation is then generated by supplying this observable as input to the t-SNE algorithm \cite{tsne}.
The output of this procedure is reported in figure \ref{fig:tsne}, and shows a large octagon consisting of mainly low-dissipation embeddings and a detached high-dissipation cluster. 
Typically, the low-dissipation events require only the $l=1$ subspace, while the rarer, high-dissipation or `bursting' snapshots have significant projections onto the $l=2$ and $l=3$ eigenspaces.
This makes it clear that there is only a single class of `bursting' event here, which is unlikely to be the case at higher Reynolds numbers.

Decoding example points from within the low-dissipation cluster reveals that its centre contains snapshots that are visually similar to the first equilibrium -- compare the middle flow field in figure 5 to figure \ref{fig:bifurcation} -- while embeddings of fields with pairs of opposite-sign vortices are situated towards the edges. 
The appearance of vortices which break the shift-reflect symmetry are indicative of secondary instabilities of the first equilibrium \cite{Lucas2014}. 
The eight sectors of the octagon-like cluster correspond to latent representations of the same recurrent patterns shift-reflected in the vertical direction. This effect is visualised in figure \ref{fig:tsne} by the square symbols in the cluster, which are eight copies of the embedding of the same vorticity field.
This simple representation of the low-dissipation dynamics is retained in all autoencoders (even $m=3$, not shown).
Low-$m$ networks do not build representations of the more complex bursting behaviour (see figure \ref{fig:loss}).

%
%
\begin{figure*}
    \centering
    \includegraphics[width=\textwidth]{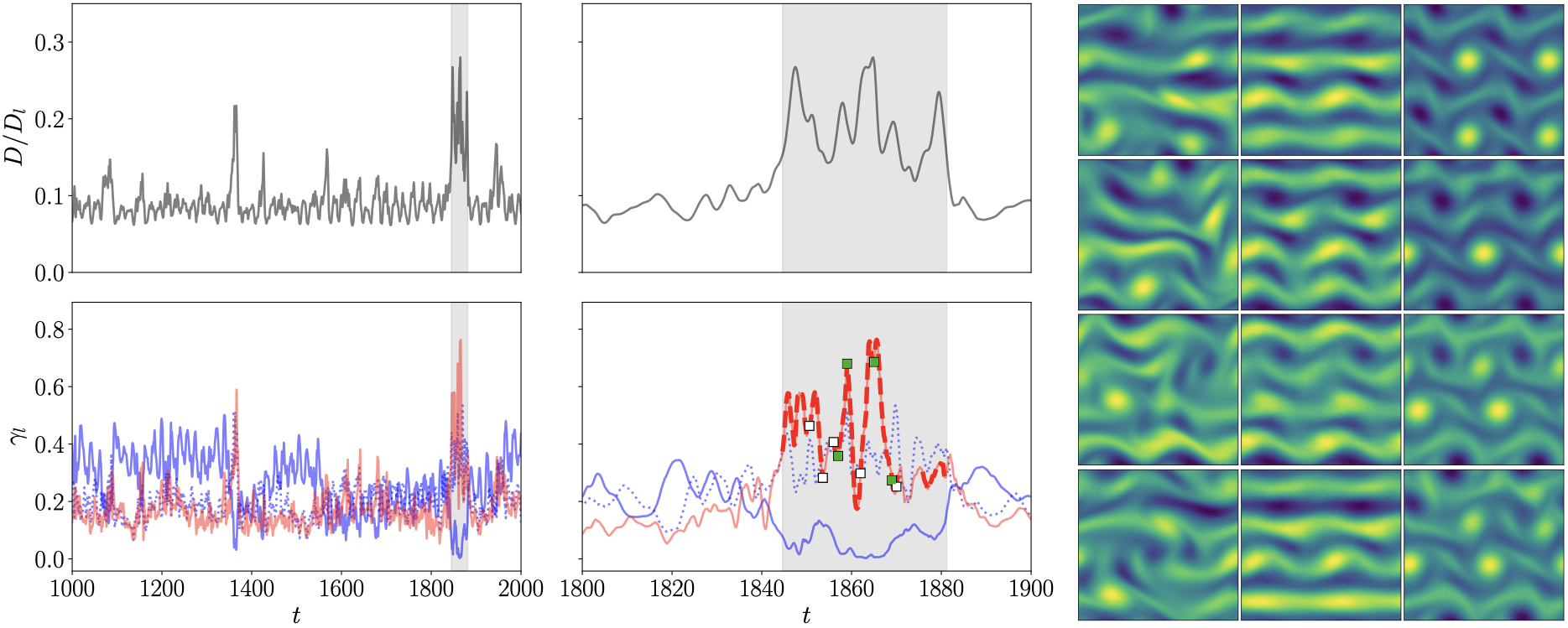}
    \caption{
        \textbf{Bursting episodes: their recurrent patterns and associated exact coherent structures.} The left and central panels show dissipation (top) and projections onto latent wavenumbers $l=1$ and $l=2$ (bottom) for a long turbulent signal. The solid blue line is the projection onto the $l=1$ mode corresponding to the primary bifurcation in the flow, $\gamma_1:=\|\pmb{\mathcal P}_0^1(\pmb{\mathscr E})\|^2$ (see text), the dashed blue line is the projection onto the \emph{rest} of the modes in the $l=1$ eigenspace , $\gamma_{1+}:=\|\sum_{j=1}^{d(1)-1}\pmb{\mathcal P}_j^1(\pmb{\mathscr E})\|^2$; the solid red line is the projection   onto the full $l=2$ eigenspace giving $\gamma_2$. The grey region is shown in more detail in the central panels, the dashed red line overlaying the $l=2$ curve identifies regions where $D/D_l > 0.15$. Square markers indicate where equilibria and travelling waves corresponding to the $l=2$ recurrent patterns were found. 
        The green markers identify the states shown on the right: for each marker we show the original snapshot, the decode of the projection onto the $l=2$ eigenspace (including the $l=0$ contribution) and the converged equilibria or travelling wave.
    }
    \label{fig:burst}
\end{figure*}
The extraction of a known equilibrium from the embeddings, and the demonstration that the flow spends much of its time nearby in phase space, highlights the advantages of autoencoders as tools for generating low-dimensional representations which can be connected to the underlying dynamics. 
More significantly, the latent Fourier decompositions also allow us to efficiently find many \emph{new} simple invariant solutions of the governing equations (\ref{eqn:NS}) in parts of phase space --the bursting events-- where current methods struggle, which we now describe.

    A long time series from a turbulent computation is examined in figure \ref{fig:burst}, visualised both in terms of dissipation and also the magnitude of the projection of the flow embeddings onto certain latent Fourier modes. 
    The `bursting' events could be classified as sections of the time series where the dissipation rate exceeds some threshold; for example $D/D_l > 0.15$ might be sensible here.
    The latent Fourier projections offer an alternative view on the bursting in terms of a distance from the simple equilibrium described above, which is central to the low-dissipation dynamics (see figure \ref{fig:tsne}).
    The solid blue line in figure \ref{fig:burst} shows the projection onto the latent Fourier mode which encodes this structure, $\pmb{\mathcal P}_0^1$. 
    As expected, this equilibrium is dominant in the embedding for the low-$D$ dynamics but becomes insignificant in the (high-$D$) bursting events.
    Bursting also exhibits a dramatic increase in importance of the $l=2$ eigenspace, as well as the other modes from within $l=1$.
    There is also a significant projection onto $l=3$ (not shown).

    Motivated by the prominent role of the $l=2$ eigenspace in the bursting, we explored how well the embedding is capturing the simple invariant solutions present in this part of phase space by supplying the decode of embeddings projected onto this space,
    \begin{equation}
        \omega^{(2)} = \mathscr D\left(\sum_{j=0}^{d(0)-1} \pmb{\mathcal P}^0_j + \left[\sum_{j=0}^{d(2)-1} \pmb{\mathcal P}^2_j\ + \text{c.c.}\right]\right)
        \label{eqn:burst_guess}
    \end{equation}
    as initial guesses in a Newton-GMRES solver searching for equilibria and travelling waves (see SI).
    Some examples of the recurrent patterns associated with this decode were included earlier in figure \ref{fig:Tspec}, these decodes have symmetry under half domain shifts $\mathscr T_{\pi}\omega^{(2)} = \omega^{(2)}$.
    This translational symmetry matches that found in the equilibrium `$E_{13}$' which was the only solution found to be important in the bursting dynamics in \cite{Farazmand2016}. 

    By constructing guesses via (\ref{eqn:burst_guess}) we have been able to converge a large number of new equilibria and travelling waves directly from the bursting snapshots themselves, as well as re-discovering $E_{13}$.   From an analysis of $\sim 20$ ``bursts'' within a time series of length $t \sim 4000$ we have found over $25$ unique solutions, usually finding at least one simple invariant solution per burst. All of our solutions have high dissipations, with the majority having values $D/D_l > 0.2$.
    We include some of these new solutions in figure \ref{fig:burst} alongside the original snapshots and the initial guess generated via equation (\ref{eqn:burst_guess}). 
    The signature of the converged solution can often be seen in  the $l=2$ recurrent pattern, which exhibits vortical features also found in the original snapshot. The fact that the solutions found seem positioned at extremes of the dissipation signal - see the middle lower plot in figure \ref{fig:burst} - is fully consistent with the picture of the turbulent trajectory  bouncing between the neighbourhoods of these unstable solutions in phase space.
    The utility of the method is that it can identify simple invariant solutions which are actually transiently visited by the dynamics {\em in the high-dimensional bursts}, which has not been possible using previous approaches.    Moreover, some of our new solutions are qualitatively different from any that have been converged before\cite{Chandler2013,Farazmand2016}, for example note the vorticity snapshots dominated by dipole structures in figure \ref{fig:new_eqs} in the SI.

\section*{Conclusion}

%
In this paper we have used deep convolutional autoencoders to construct efficient low-dimensional representations of monochromatically forced, two-dimensional turbulence at $Re=40$.
The networks are highly effective at identifying recurrent spatial patterns in the vorticity field -- common wavenumber pathways in a Fourier representation -- in striking contrast with standard dimensionality reduction techniques.
By exploiting a continuous symmetry we have developed an {\sl interpretable} latent Fourier decomposition of the embeddings: the latent Fourier modes can be decoded into physically meaningful fields. This has allowed us to reveal the structure of state space underlying the dynamics. One equilibrium (the primary bifurcation) dominates the quiescent low-dissipation dynamics while one grouping of simple invariant solutions organise a single type of high dissipation bursting event which occurs intermittently. The success of latent Fourier analysis in identifying dynamically important solutions of high dissipation for the bursting episodes is particularly noteworthy as previous methods \cite{Chandler2013,Farazmand2016} have struggled to do this. 
Going forward, these new solutions present a way of charting the bursting dynamics, as latent Fourier decompositions provide us with a natural metric for measuring which solution a turbulent orbit is nearest to.
Moreover, latent Fourier analysis also allows us to efficiently find large numbers of periodic orbits, including those in previously unreachable parts of the state space (see the example in the SI), than has been previously possible\cite{PageAPS2018} and we plan to report the results of these searches in the near future.


The results presented here clearly show that harnessing machine learning techniques to allow the building blocks of a flow representation to \emph{design} themselves based on the flow dynamics is a significant step forward.  The blocks which emerge are the principal spatial patterns or coherent structures observed in the flow and, intriguingly, can accurately capture simple invariant solutions embedded in the turbulent attractor without the solutions ever being realised precisely. 
This opens up the possibility of an easily automated, direct approach for both identifying when the flow is in the neighbourhood of a state in phase space and evaluating the probability of being there. Turbulent statistics could then be predicted through a weighted sum over relevant states, for example, in the spirit of periodic orbit theory \cite{Artuso1990a,Artuso1990b}. However, in the immediate future, the natural next step is to apply these techniques at much higher Reynolds numbers and in three dimensions. In these extensions, assessing how much data is needed to power this approach will also be an important consideration.


\section*{Acknowledgements}
    JP acknowledges support from the Sultan Qaboos Fellowship at Corpus Christi College, University of Cambridge.
    MPB acknowledges support from the Simons Foundation and also from NSF Division of Mathematical Sciences grant DMS1715477.
\section*{Author Contributions} 
    The project was conceived and developed by all three authors who also contributed significantly to the writeup. 
    JP did the majority of the computations with help from MPB.
\section*{Competing Interests} The authors declare that they have no
competing financial interests.
\section*{Correspondence} Correspondence and requests for materials
should be addressed to J. Page~(email: Jacob.Page\@ ed.ac.uk).

%

\pagebreak
\begin{center}
\textbf{\large Supplementary Information: Revealing the state space of turbulence using machine learning}
\end{center}
\beginsupplement

\section{Data}
Training data are generated by solving equation (\ref{eqn:NS}) at fixed $Re=40$. 
For spatial discretisation we apply a two-dimensional Fourier transform at a resolution $N_x\times N_y = 128\times 128$; de-aliasing is applied according to the $2/3$-rule. 
For timestepping, an implicit Crank-Nicholson scheme is employed for the diffusion term and Heun's method is used for the nonlinear advection terms. For further details see \cite{Chandler2013, Lucas2014}. 

The training dataset is constructed from $1000$ independent trajectories each consisting of $100$ snapshots separated by $\Delta t=0.5$. 
Each trajectory was generated by simulating (\ref{eqn:NS}) from a randomly perturbed initial condition and discarding an initial transient. 
The vorticity fields are all normalized, $\omega_{\text{train}} = \omega / \omega_M$, where $\omega_M=15$, which ensures $|\omega_{\text{train}}|<1$.
    Each vorticity field then has a random symmetry transform applied $\omega \to \mathscr T_{s}\mathscr S^k \mathscr R^j \omega$ to ensure the network sees the full state space.

The test dataset used to generate the figures in this paper is constructed from a further $1000$ trajectories in the same way.  

\section{Autoencoder architecture}
    The autoencoders discussed in this paper were all implemented using the \texttt{Keras} library
 \cite{chollet2015}.
    All share a common structure consisting of a series of five convolutional layers with periodic padding.
    The number of filters (and kernel size) decreases sequentially, $128(8,8)\to 64(8,8)\to 32(4,4)\to 16(4,4)\to 8(2,2)$.
    Nonlinear ReLU activation is applied to the output of each layer.
    Each convolution is followed by a max pooling operation over regions of size $2 \times 2$.
    
    The convolutions are followed by three fully connected layers, all with ReLU activation, which run $128\to m \to 128$.
    The fully connected layers are followed by a series of five convolutional layers which mimics the encoder described above, with up sampling applied after each convolution on patches $2\times 2$.
    A final convolutional layer with a $\tanh$ activation produces the output.
    
    We trained networks with embedding dimension $m \in \{3,8,16,32,48,64,96,128\}$.
    Training was performed for $800$ epochs for batch sizes of $64$ and $128$ using an Adam optimizer with a learning rate of $0.001$ or $0.0003$.
    The results presented in this paper were generated using the best performing model at a given $m$.
    In order of increasing $m$ these hyper parameters are ($m$, learning rate, batch size): 
    (3, 0.001, 64);
    (8, 0.001, 128);
    (16, 0.001, 128);
    (32, 0.001, 128);
    (48, 0.0003, 64);
    (64, 0.0003, 128);
    (96, 0.0003, 64);
    (128, 0.0003, 128).

\section{Details on symmetry operators}
\begin{figure*}
    \centering
    \includegraphics[width=0.25\textwidth]{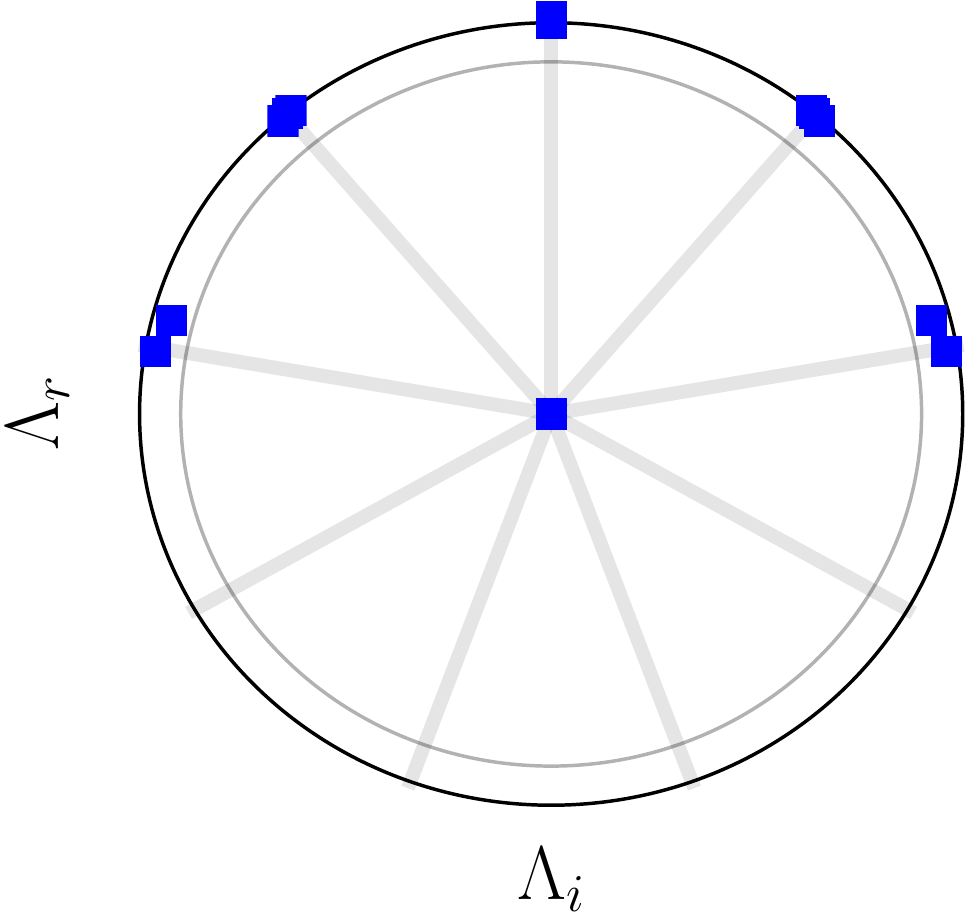}
    \includegraphics[width=0.25\textwidth]{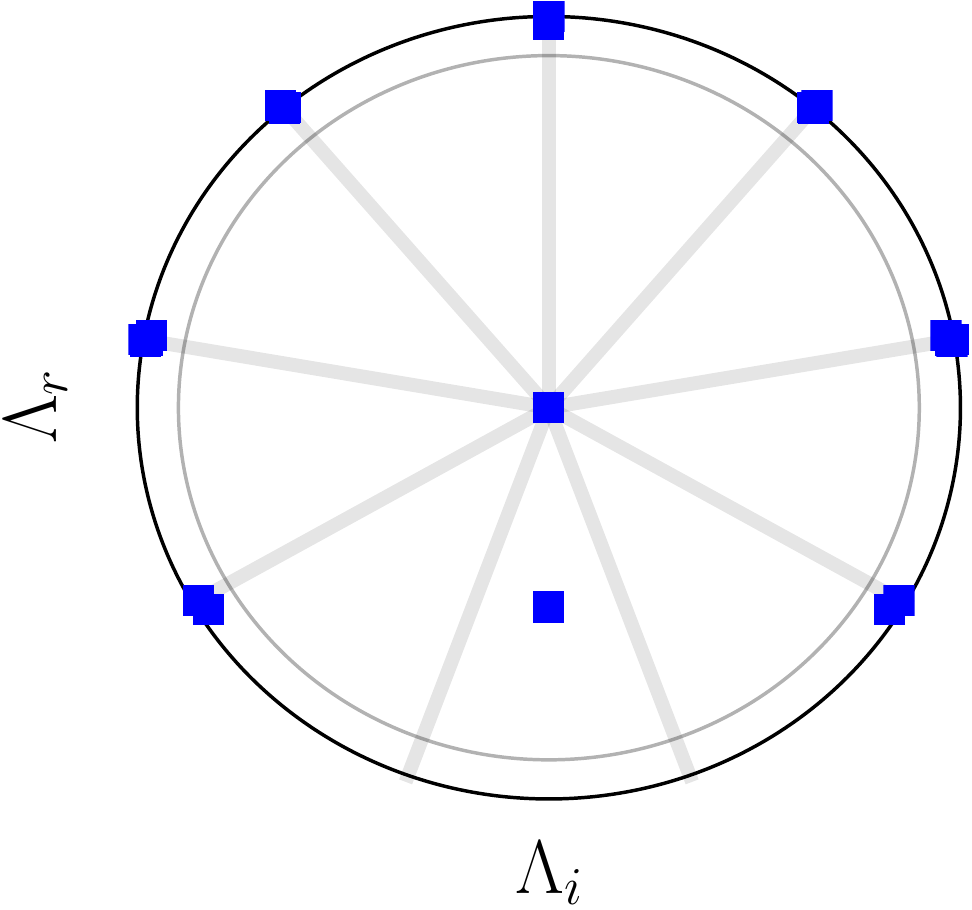}
    \includegraphics[width=0.25\textwidth]{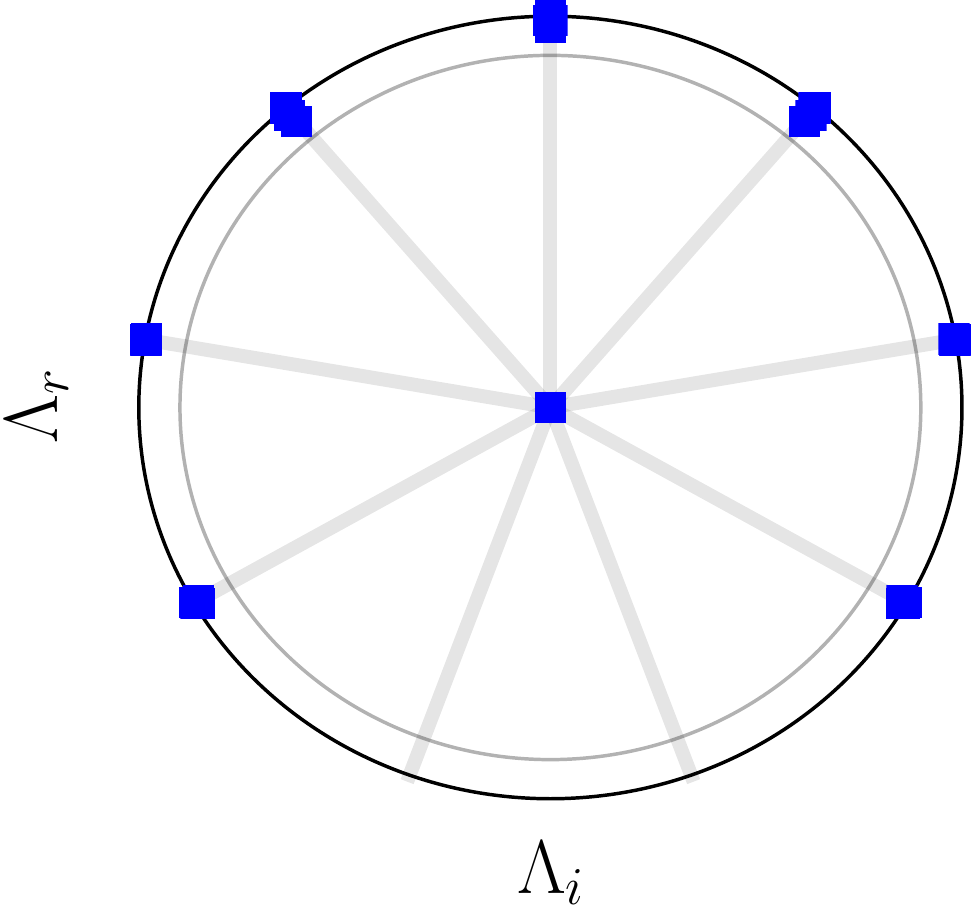}
    \\
    \includegraphics[width=0.25\textwidth]{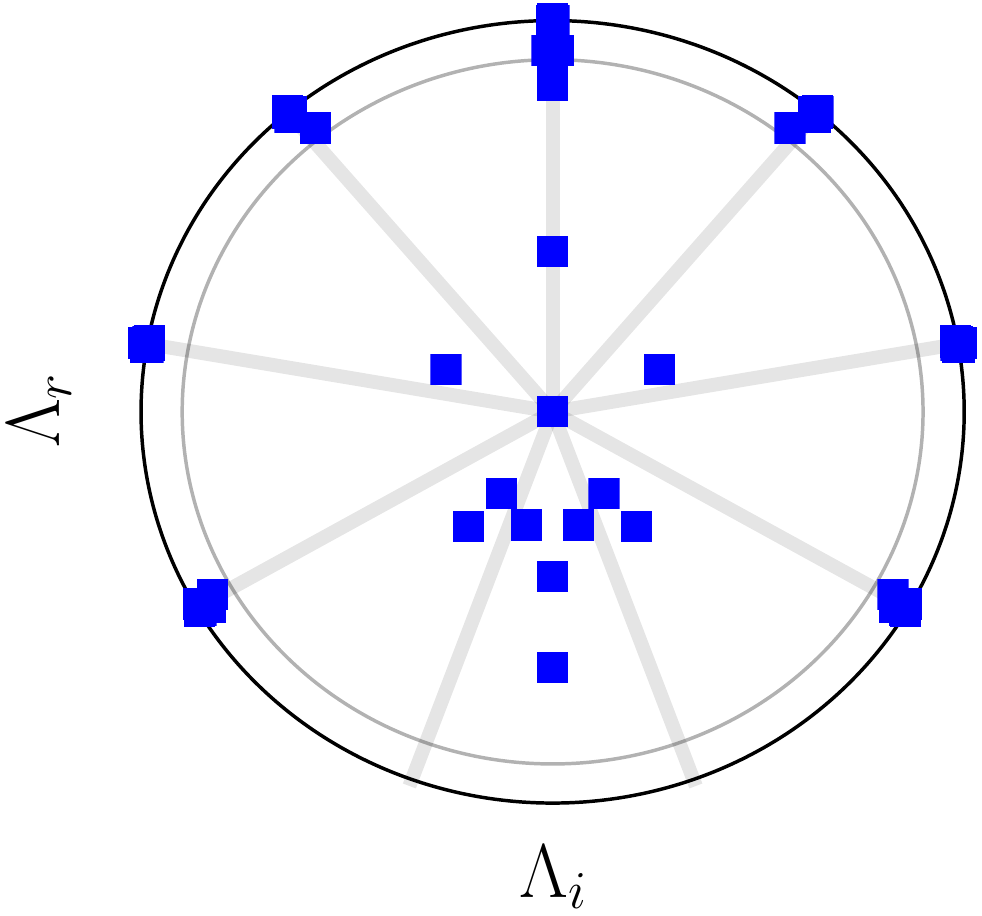}
    \includegraphics[width=0.25\textwidth]{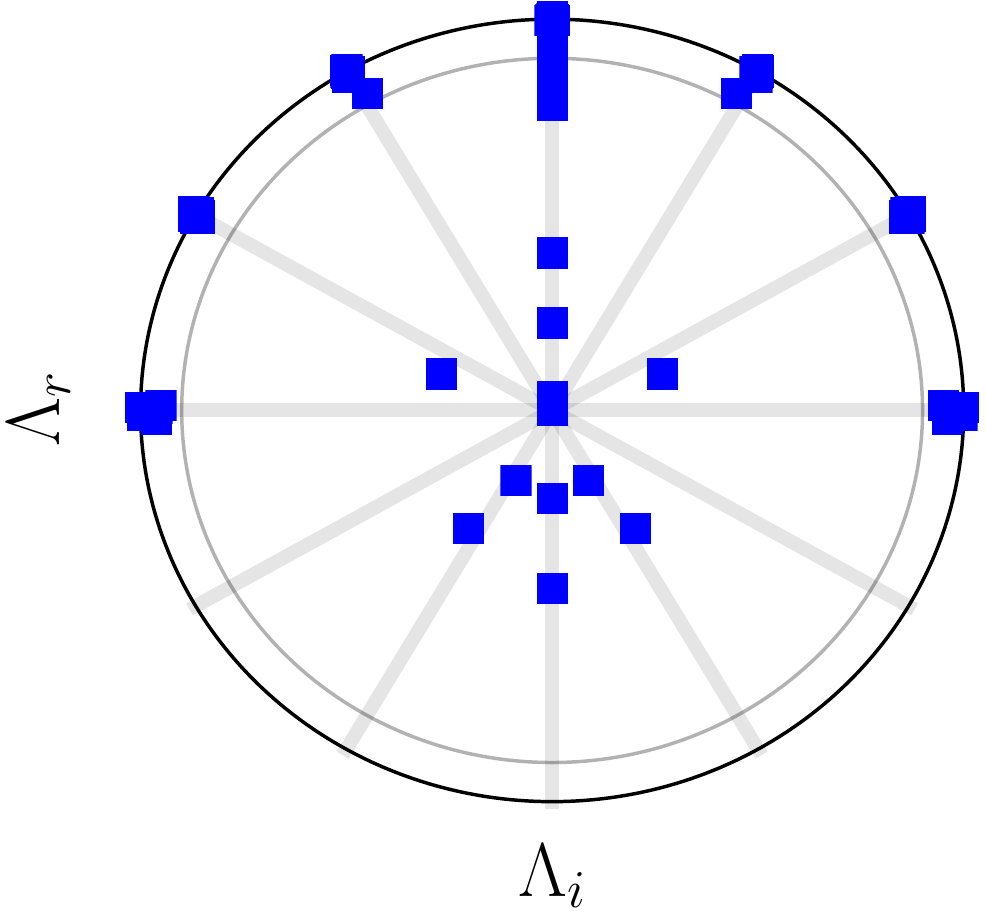}
    \includegraphics[width=0.25\textwidth]{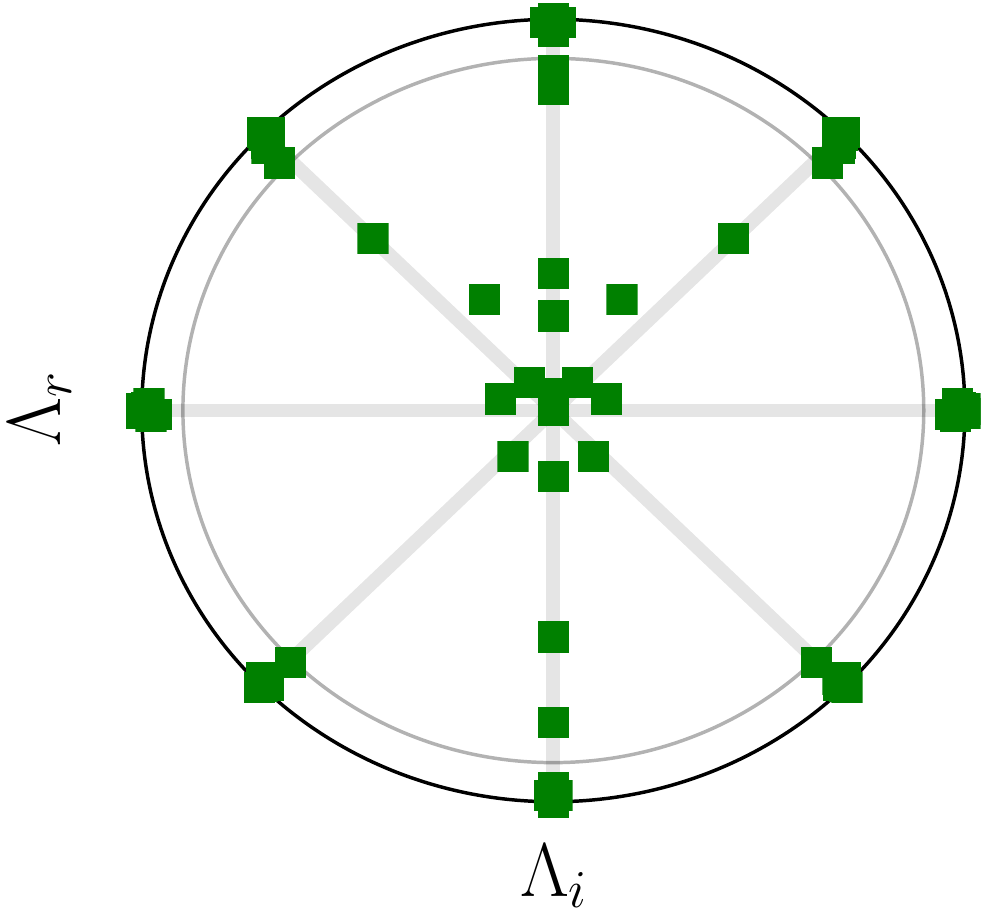}
    \caption{\textbf{Eigenvalue spectra of the latent symmetry operators for various embedding sizes $m$ and shifts $\alpha$.} 
    (Top) Fixed $\alpha=2\pi/9$; from left to right $m=16,32,64$.
    (Bottom) Fixed $m=96$. Left panel has $\alpha=2\pi/9$, central panel $\alpha=2\pi/12$. 
    The final panel (green spectrum) shows eigenvalues of a numerical approximation to the vertical shift-reflect operator, $\mathbf S$.
    In all cases, the outer black circle is $|\Lambda| = 1$, the grey interior circle is $|\Lambda|=0.9$. The rays from the origin identify the relevant roots of unity.}
    \label{fig:Tspec_methods}
\end{figure*}
    We construct shift operators $\mathbf T_{\alpha}$ for each network by first assembling a matrix of embeddings, 
$\mathbf E:=\begin{bmatrix}\pmb{\mathscr E}(\omega_1) & \cdots & \pmb{\mathscr E}(\omega_N)\end{bmatrix}$,
along with another data matrix built from embeddings of the same vorticity fields shifted by $\alpha$ in $x$,
$\mathbf E':=\begin{bmatrix}\pmb{\mathscr E}(\mathscr T_{\alpha}\omega_1) & \cdots & \pmb{\mathscr E}(\mathscr T_{\alpha}\omega_N)\end{bmatrix}$ 
An approximate shift operator is then determined from a least-squares fit over the test set, $\mathbf T_{\alpha} = \mathbf E' \mathbf E^{+}$, where $\mathbf E^+$ is the Moore-Penrose pseudo inverse of $\mathbf E$. 
This algorithm is well-known in the fluid dynamics community, where it is typically applied to temporally-spaced flow snapshots to extract `dynamic modes' with an exponential dependence on time \cite{Schmid2010, Rowley2009, Page2019}.

    We compute a set of several shift operators in latent space, $\mathbf T_{\alpha}$, for various network dimensions $m$ and shifts $\alpha$. 
    The eigenvalue spectra of some of these operators are reported in figure \ref{fig:Tspec_methods} in `timestepper' form. 
    Latent wavenumbers can be extracted via $l = \log \Lambda / (i\alpha)$. 
    At fixed $\alpha$, the number of required latent wavenumbers saturates at $l=3$ beyond $m=32$.
    At fixed $m$, no further latent wavenumbers are recovered for shifts $\alpha < \pi/6$. 
    Therefore, any shift $\alpha < \pi/6$ is sufficient to perform a latent Fourier transform without any aliasing issues.
    
    The final spectrum reported in figure \ref{fig:Tspec_methods} (green symbols) corresponds to an operator that performs shift-reflect operations in the latent space,
    \begin{equation}
        \mathbf S \pmb{\mathscr E}(\omega) = \pmb{\mathscr E}(\mathscr S\omega).
    \end{equation}
    The eigenvalues approximate the eight eighth roots of unity, as expected.

    As discussed in the body of the paper, the latent wavenumbers $\{l\}$ are degenerate. 
    Therefore, we are free to choose a basis with each eigenspace. 
    We first construct an arbitrary bi-orthogonal basis from the numerically computed left- and right-eigenvectors of $\mathbf T_{\alpha}$,
    \begin{align}
        \boldsymbol \Xi_l &= \mathbf V_l \mathbf R^{-1}, \\
        \boldsymbol \Xi^{\dagger}_l &= (\mathbf Q^{-1}\mathbf W^H_l)^H, 
    \end{align}
    where the columns of $\mathbf V_l$ are the numerically computed right eigenvectors of $\mathbf T_{\alpha}$ corresponding to wavenumber $l$; the columns of $\mathbf W_l$ are the left eigenvectors.
    The matrices $\mathbf Q$ and $\mathbf R$ are the QR decomposition of $\mathbf W^H_l\mathbf V_l$.
    The columns of $\boldsymbol \Xi$ and $\boldsymbol \Xi^{\dagger}$ form a biorthogonal basis, $\boldsymbol \xi_i^{\dagger H} \boldsymbol \xi_j = \delta_{ij}$.
    We then compute the projections of our test set of embeddings within each eigenspace, and perform PCA on the resulting data matrix to form an orthogonal basis ordered by `energy'.

\begin{figure*}
    \centering
    \includegraphics[width=0.7\textwidth]{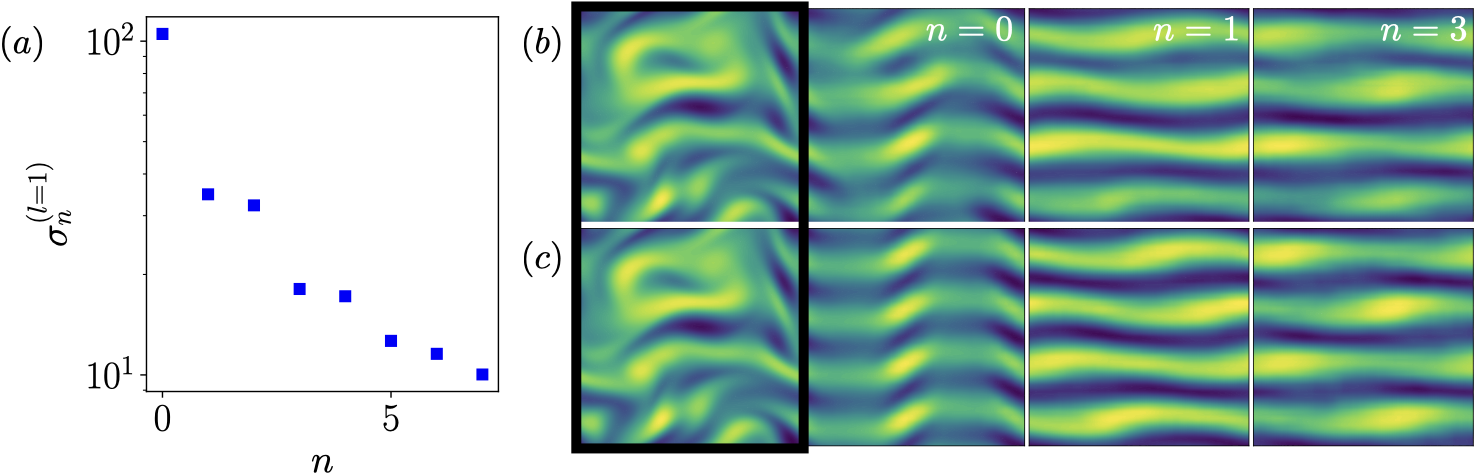}
    \caption{\textbf{Visualisation of the degenerate eigenspace $l=1$ via decodes of projections onto its PCA modes for an example snapshot.}
    (a) Singular values from the $l=1$ subspace.
    (b) An example vortical snapshot (left panel, highlighted with a black box) and the decode of the projection of its embedding onto \emph{individual} PCA modes $\{\boldsymbol u^{(1)}_n\}$ within the $l=1$ subspace, note
    the $l=0$ projection is also included (see equation \ref{eqn:l_pca_v1} in the text).
    (c) As (b) but only the first PCA mode from the $l=0$ subspace is used (see equation \ref{eqn:l_pca_v2} in the text).}
    \label{fig:lat_pca}
\end{figure*}
    As described in the body of the paper, visualisation of individual PCA modes associated with a particular latent wavenumber $l\geq 1$ also requires some contribution for the $l=0$ eigenspace to be included in the decode. 
    In the text we used only the leading PCA mode from this space (equation \ref{eqn:l_pca_v2}), though other choices are possible.
    For example, the full $l=0$ subspace can be used,
    \begin{equation}
        \omega_n^{(1)} = \mathscr D\left(\sum_{j=0}^{d(0)-1} \pmb{\mathcal P}^0_j + [\pmb{\mathcal P}_n^1 + \text{c.c.}]\right).
        \label{eqn:l_pca_v1}
    \end{equation}
    An example is included in figure \ref{fig:lat_pca}, where we report the singular values in the $l=1$ eigenspace and the decodes of projections onto individual PCA modes for an arbitrary snapshot using both equation (\ref{eqn:l_pca_v2}) and equation (\ref{eqn:l_pca_v1}).
    As noted in the main text, the leading PCA mode in $l=1$ is revealed to be visually similar to the primary bifurcation of the flow at lower $Re$. 
    Higher order PCA modes also have a large degree of discrete symmetry.
    For example, the slanted vortical structures associated with mode $\boldsymbol u_1^{(1)}$ are symmetric under rotation and whole-wavelength shifts,
$\overline{\omega}_0^{(1)} = \mathscr R \,  \overline{\omega}_0^{(1)}$, $\overline{\omega}_0^{(1)}=\mathscr S^{2m}\, \overline{\omega}_0^{(1)}$. 
The projection onto $\boldsymbol u_2^{(1)}$ (not shown) is simply the shift-reflect of $\boldsymbol u_1^{(1)}$, $\boldsymbol u_2^{(1)} = \mathscr S^{2m+1} \boldsymbol u_1^{(1)}$.
The same set of symmetries hold for the vortex blobs obtained when decoding projections onto $\boldsymbol u_3^{(1)}$ and $\boldsymbol u_4^{(1)}$.

\section{Simple invariant solutions}
\begin{figure*}
    \centering
    \includegraphics[width=0.15\textwidth]{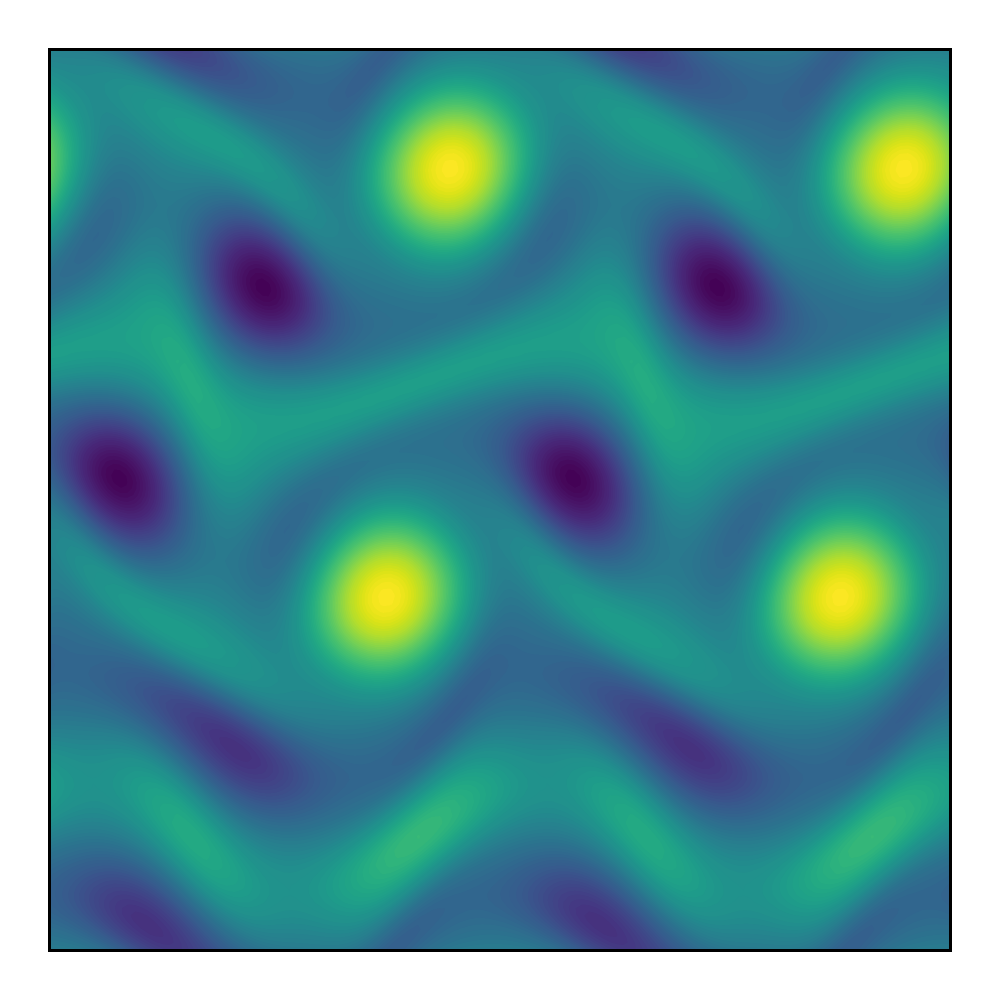}
    \includegraphics[width=0.15\textwidth]{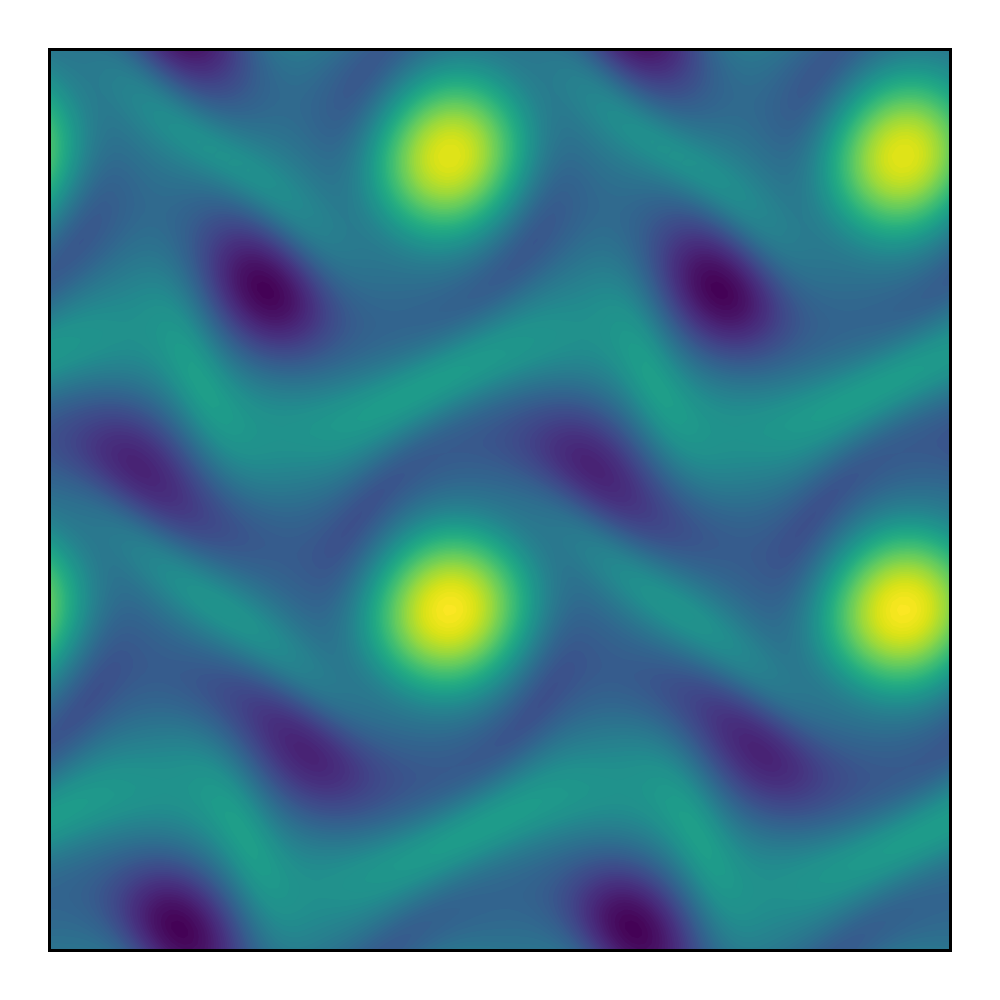}
    \includegraphics[width=0.15\textwidth]{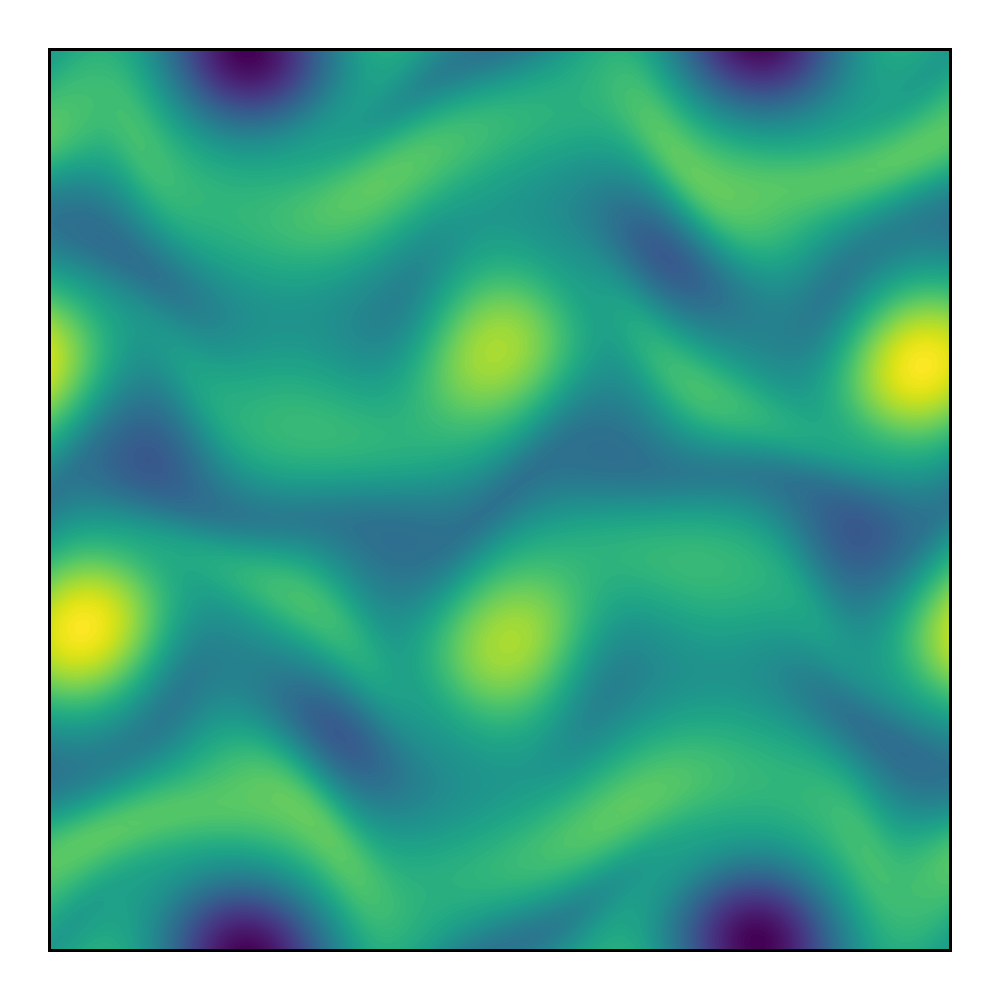}
    \includegraphics[width=0.15\textwidth]{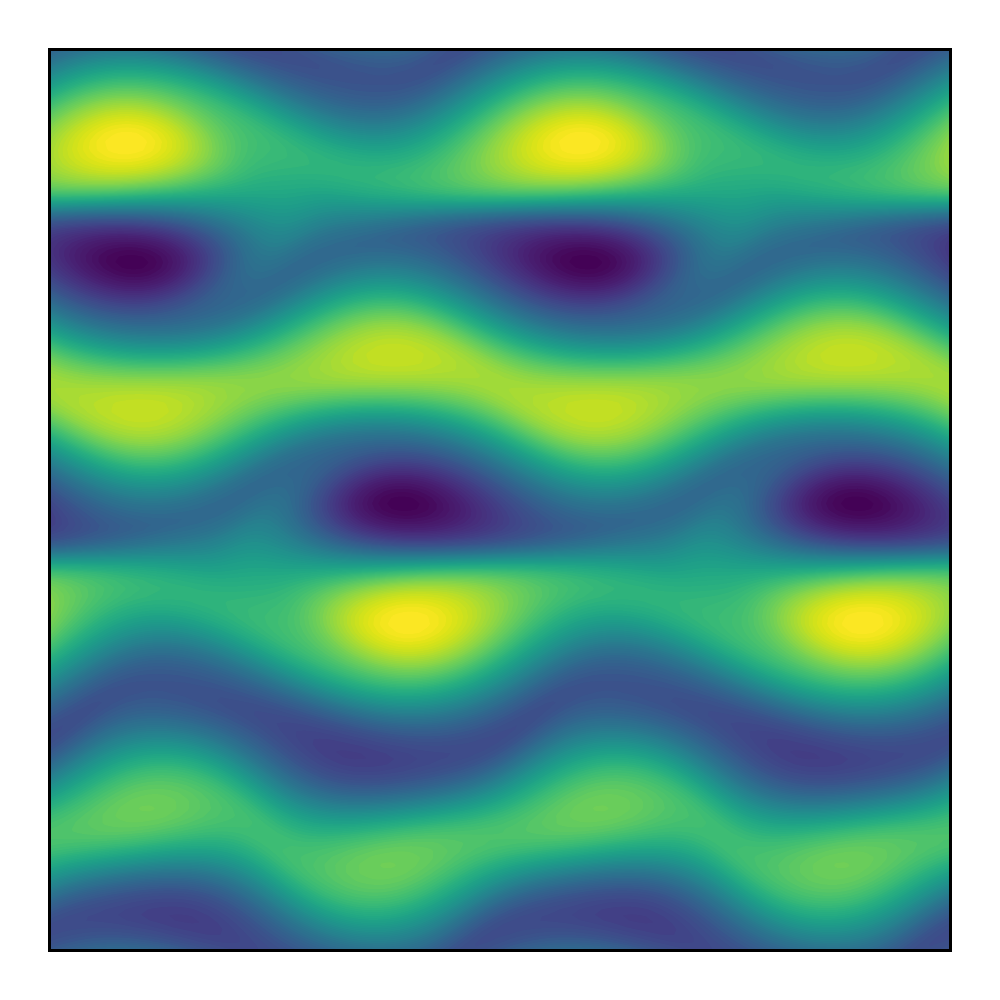}
    \\
    \includegraphics[width=0.15\textwidth]{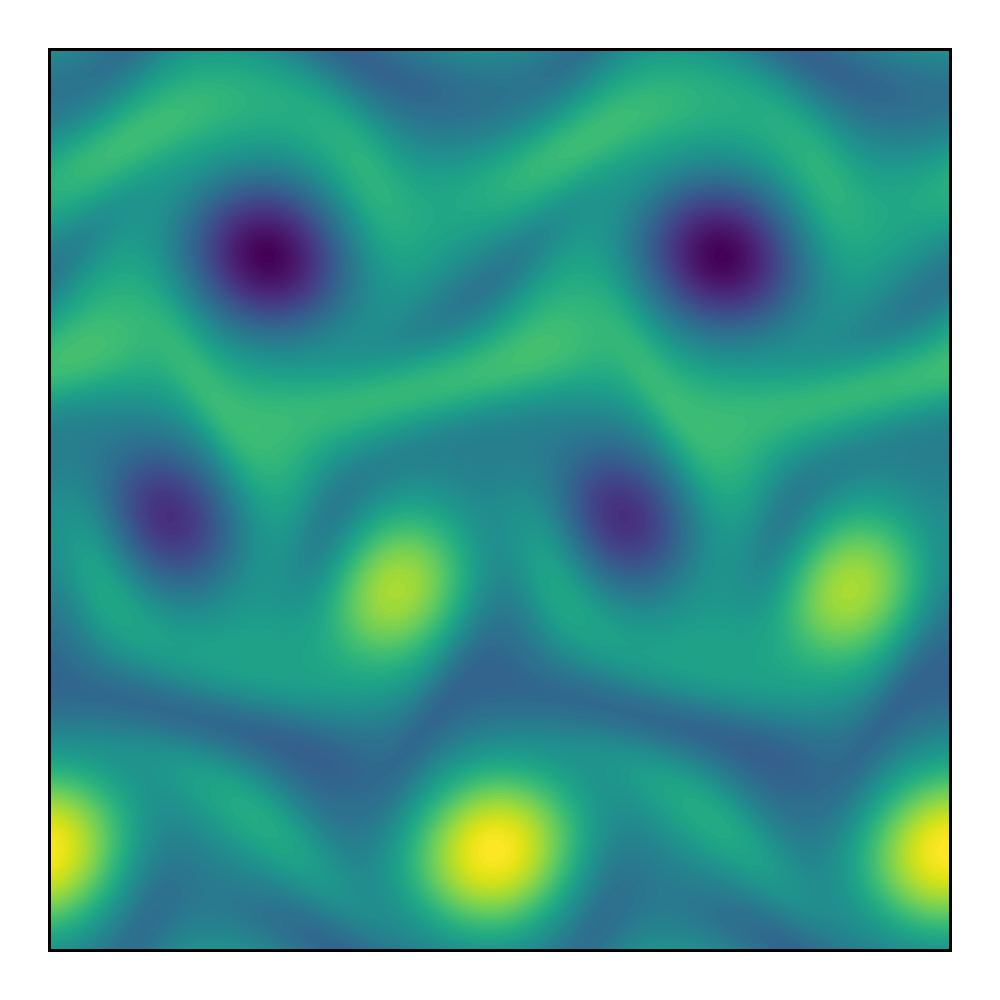}
    \includegraphics[width=0.15\textwidth]{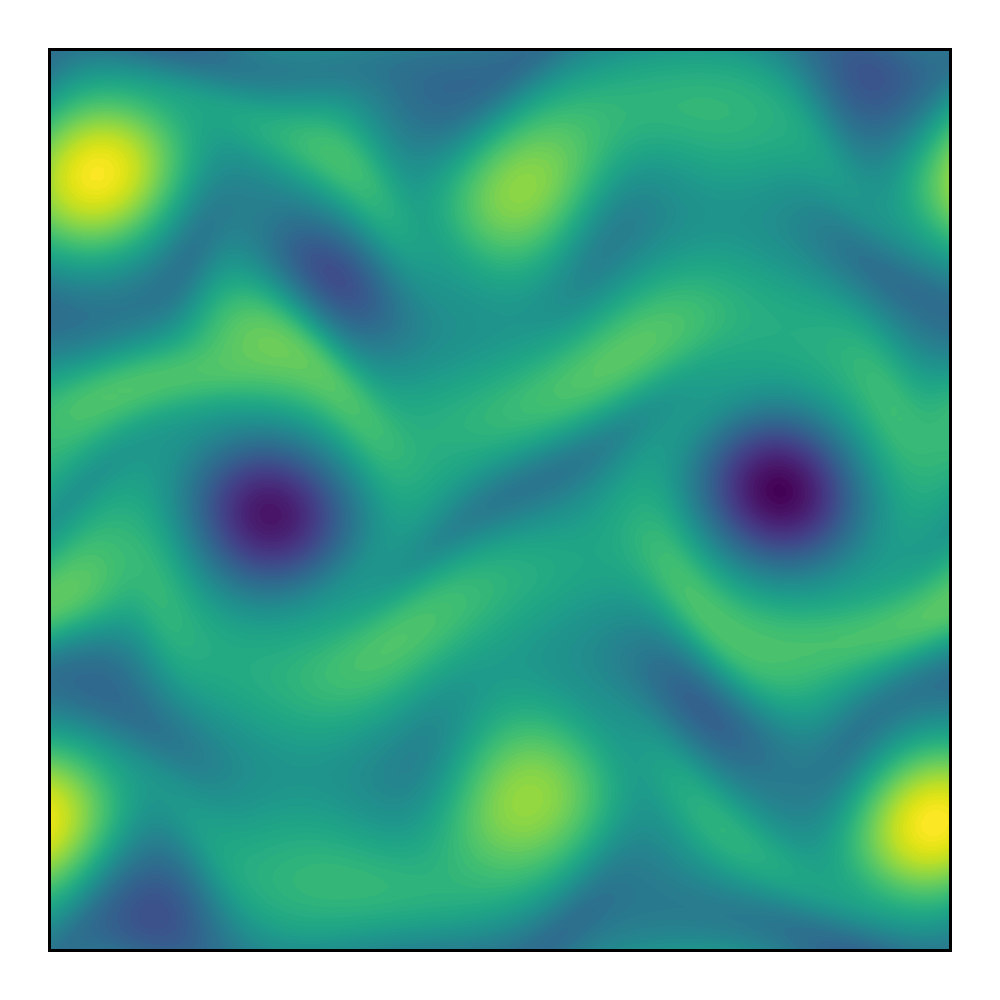}
    \includegraphics[width=0.15\textwidth]{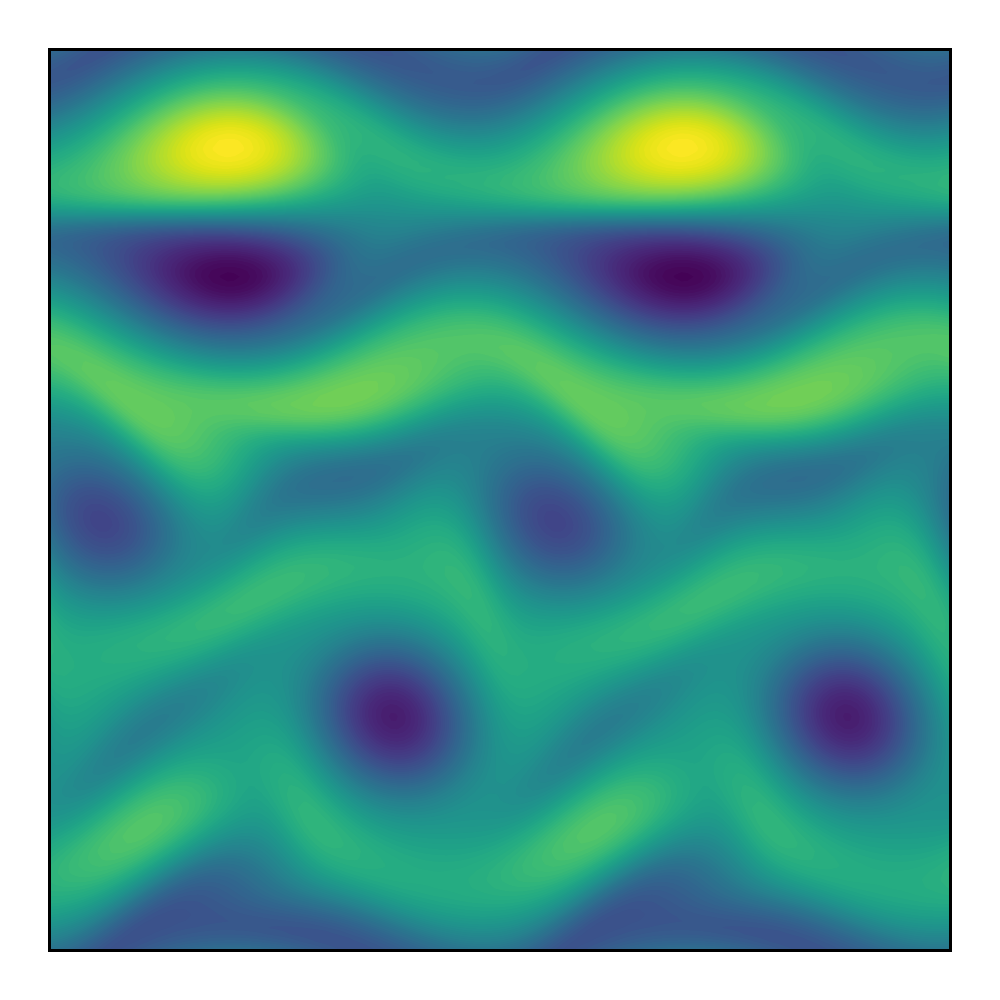}
    \includegraphics[width=0.15\textwidth]{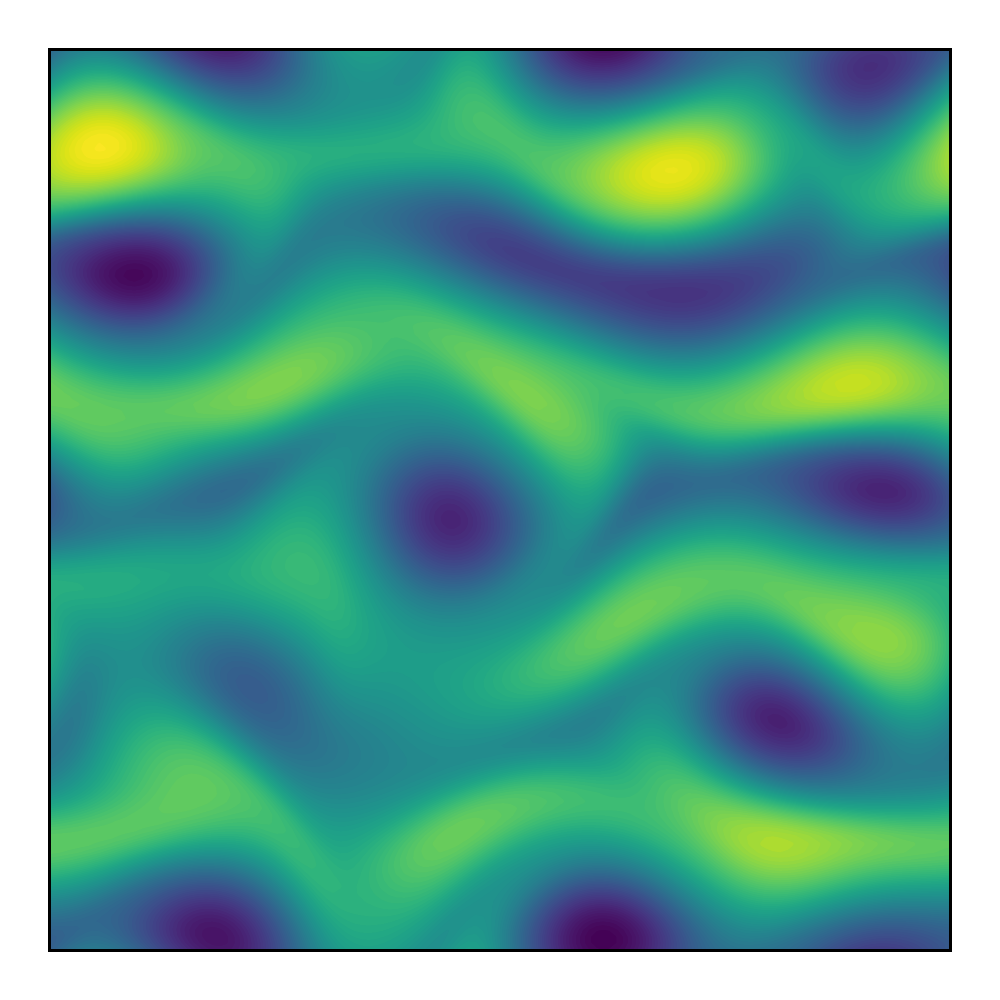}
    \caption{\textbf{Some new bursting equilibria and travelling waves.} (Top) Vorticity field for four new equilibria converged from projections and decodes onto $l=2$. From left to right the dissipation rate of these new solutions is $D/D_l=0.1974, 0.2042, 0.2233, 0.6058$. (Bottom) New travelling waves. From left to right the phase speed of these structures is $c=0.0006,0.0053,0.3178,0.1059$ and the dissipation rate is $D/D_l=0.2184,0.2257,0.4357,0.4536$.}
    \label{fig:new_eqs}
\end{figure*}
\begin{figure*}
    \centering
    \includegraphics[width=\textwidth]{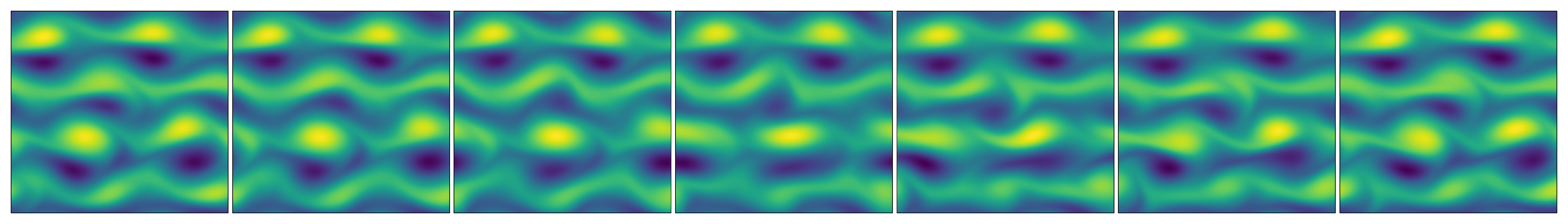}
    \caption{\textbf{Bursting periodic orbit.}
    Snapshots separated by $\delta t =0.5$ from a newly-discovered (relative) periodic orbit with period $T=3.41276$ and shift $s=-0.56022 $.}
    \label{fig:new_upo}
\end{figure*}
    Equilibria and travelling waves correspond to solutions of (\ref{eqn:NS}) satisfying
    \begin{equation}
        \mathscr T_{s=cT} \omega(\mathbf x, t+T) = \omega(\mathbf x, t) \quad \forall T
    \end{equation}
    where the shift $s=cT$ is set by the fixed wavespeed, and $c=0$ for a pure equilibrium.
    The guesses take the form $(\omega_g, s_g)$, where we set the shift $s_g=c_g T_{int} = 0$ and the integration time is held fixed at $T_{int}=3$.   
    The vorticity guess is the decode of a projection onto the $l=2$ eigenspace (see equation (\ref{eqn:burst_guess}) in the main text) from the embedding of a snapshot from within a `bursting' episode. 

    The guesses are input into a Newton-Raphson algorithm which has been described extensively in previous research \cite{Viswanath2007,Gibson2008,Chandler2013}.
    The size of the Jacobian matrix makes direct computation prohibitively expensive, and updates to the solution $(\delta \omega, \delta s)$ are instead computed within a Krylov subspace (Newton-GMRES) which requires computation only of the action of the Jacobian on a vector.
    A hookstep is used to constrain updates of the guess to within a specified trust region 
 \cite{Viswanath2007}.

    Some example solutions converged from $l=2$ projections within bursting events are displayed in figure \ref{fig:new_eqs}.
    Note the dipole structures seen in the simple invariant solutions with the highest dissipation values have not been seen in previously discovered exact coherent structures \cite{Chandler2013,Farazmand2016}.
    We have also converged a relative periodic orbit from within a bursting episode and have included snapshots from the evolution of this solution in figure \ref{fig:new_upo}.
    The flow field is dominated by four dipole structures; the lower pair propagate through the domain while the upper pair remain fixed in place. 
    This bursting periodic orbit is qualitatively different from any documented previously \cite{Chandler2013,Lucas2014}.

%


\end{document}